# Automated Pipeline for EEG Artifact Reduction (APPEAR) Recorded during fMRI

Ahmad Mayeli[1,2*], Obada Al Zoubi[1,2*], Kaylee Henry[1,3], Chung Ki Wong[1], Evan J. White[1], Qingfei Luo[1], Vadim Zotev[1], Hazem Refai[2], Tulsa 1000 Investigators[1#], Jerzy Bodurka[1,4**]

[1]Laureate Institute for Brain Research, Tulsa, OK, United States
[2]Electrical and Computer Engineering, University of Oklahoma, Tulsa, OK, United States
[3]Department of Biomedical Engineering, University of Arkansas, Fayetteville, AR, United States
[4]Stephenson School of Biomedical Engineering, University of Oklahoma, Tulsa, OK, United States

*** Co-first Authors**
Ahmad Mayeli and Obada Al Zoubi

**** Correspondence:**
Jerzy Bodurka
jbodurka@laureateinstitute.org

[#]The Tulsa 1000 Investigators include the following contributors: Robin Aupperle, Ph.D., Jerzy Bodurka, Ph.D., Justin Feinstein, Ph.D., Sahib S. Khalsa, M.D., Ph.D., Rayus Kuplicki, Ph.D., Martin P. Paulus, M.D., Jonathan Savitz, Ph.D., Jennifer Stewart, Ph.D., Teresa A. Victor, Ph.D.

**Acknowledgments**

We are particularly grateful to Julie Owen, Greg Hammond, Bill Alden, and Julie DiCarlo for helping with MRI and EEG-fMRI scanning. We would like to thank Dr. Brett Bays of Brain Visions, LLC, for their help and technical support. This work was supported in part by W81XWH-12-1-0697 award from the U.S. Department of Defense (CKW, VZ, JB) National Institute of General Medical Sciences, National Institutes of Health 1P20GM121312 award (JB, T1000), the Laureate Institute for Brain Research (LIBR), and the William K. Warren Foundation.

## Abstract

*Objective.* Simultaneous EEG-fMRI recordings offer a high spatiotemporal resolution approach to study human brain and understand the underlying mechanisms mediating cognitive and behavioral processes. However, the high susceptibility of EEG to MRI-induced artifacts hinders a broad adaptation of this approach. More specifically, EEG data collected during fMRI acquisition are contaminated with MRI gradients and ballistocardiogram (BCG) artifacts, in addition to artifacts of physiological origin. There have been several attempts for reducing these artifacts with manual and time-consuming pre-processing, which may result in biasing EEG data due to variations in selecting steps order, parameters, and classification of artifactual independent components. Thus, there is a strong urge to develop a fully automatic and comprehensive pipeline for reducing all major EEG artifacts. In this work, we introduced an open-access toolbox with a fully automatic pipeline for reducing artifacts from EEG data collected simultaneously with fMRI (refer to APPEAR). *Approach.* The pipeline integrates average template subtraction and independent component analysis (ICA) to suppress both MRI-related and physiological artifacts. To validate our results, we tested APPEAR on EEG data recorded from healthy control subjects during resting-state ($n$=48) and task-based (i.e., event-related-potentials [ERP]; $n$=8) paradigms. The chosen gold standard is an expert manual review of the EEG database. *Main results.* We compared manually and automated corrected EEG data during resting-state using frequency analysis and continuous wavelet transformation and found no significant differences between the two corrections. A comparison between ERP data recorded during a so-called stop-signal task (e.g., amplitude measures and signal-to-noise ratio) also showed no differences between the manually and fully automatic fMRI-EEG-corrected data. *Significance:* APPEAR offers the first comprehensive open-source toolbox that can speed up advancement of EEG analysis and enhance replication by avoiding experimenters' preferences while allowing for processing large EEG-fMRI cohorts composed of hundreds of subjects with manageable researcher time and effort.

## 1. Introduction

Electroencephalography (EEG) and functional Magnetic Resonance Imaging (fMRI) have both been widely used as noninvasive and safe techniques for detecting and characterizing changes in brain states and their relation to neuronal activity (1). Simultaneous EEG-fMRI leverages fMRI's capacity to measure whole brain hemodynamic activities at the high spatial resolution and high temporal resolution of EEG signals, directly reflecting electrophysiological brain activities (2). Furthermore, EEG is a direct measure of brain activity, while fMRI is an indirect measure; therefore, combining these modalities aids in validation and offers a more comprehensive understanding of spatial and temporal activities in the brain (3). However, obtaining high-quality EEG data from simultaneous EEG-fMRI experiments is difficult and faces several technical challenges (4). Recording EEG inside an MRI scanner and during fMRI acquisition results in EEG signal contamination from MRI-related artifacts. The MRI gradient-induced artifact (gradient artifact) results from a combination of switching magnetic field gradients required for spatial encoding during the fMRI acquisition. The ballistocardiogram (BCG) artifact appears to be a result of cardiac activity-induced head movements in the static polarizing $B_0$ magnetic field inside the MRI scanner (5). Other types of artifacts, such as muscle and ocular artifacts, can be present in EEG data regardless of whether the EEG is recorded inside or outside the MRI scanner (6, 7).

After years of developing simultaneous EEG-fMRI techniques, several methods have been proposed for reducing artifacts from EEG data based on three main strategies. First, the artifact reduction strategy employs templates from BCG and gradient (i.e., MRI-related) artifacts that are subtracted from the main signal (2, 8-10). To date, the average artifact subtraction method (AAS) (9, 10) is one of the most common approaches in reducing BCG - and especially gradient artifacts. The AAS method uses the repetitive pattern of the gradient and BCG artifacts to generate an artifact template by averaging the EEG intervals that are contaminated by the artifact to then subtract from the EEG signal. Though the AAS can effectively reduce BCG and gradient artifacts, some residual artifacts remain when this method is applied to raw EEG data in both real-time and offline (2). To get the best results, this method requires a high reproducibility of the artifact's pattern, shape, and duration, which depend on the MRI scanner's hardware quality, to generate highly reproducible gradient waveforms and excellent time synchronization between the MRI and EEG data acquisition systems. With modern MRI hardware, these requirements can be achieved, and AAS typically provides excellent gradient artifact suppression (11, 12). However, using AAS for reducing BCG artifacts requires additional consideration due to the artifact's inherent variability. In other words, AAS assumes that BCG artifacts are relatively stable over time; however, this is not always the case. Due to subjects' physiological variabilities, BCG artifacts are known to fluctuate over time, resulting in excessive residual BCG artifacts when using AAS. (2) suggested a more comprehensive approach based on AAS, namely the optimal basis set (OBS), for reducing MRI-related artifacts. To minimize the effect of residual gradient and BCG artifacts, principal component analysis (PCA) was proposed to capture the temporal variations of BCG artifacts and then regress them out from EEG data. A recent study proposed modelling the gradient artifact directly using the known MRI sequence gradient waveforms in order to reduce motion-affected gradient artifacts (13). Second, another common artifact reduction approach is employing an extra sensor during simultaneous EEG-fMRI recording for capturing MRI-related artifacts and further subtracting them from the raw data (14-20). For instance, (14) utilized a piezoelectric motion sensor to estimate motion and BCG artifacts. They calculated the correlation between the motion sensor signal and EEG signal to further design a Kalman filter to remove BCG artifacts. (17) introduced a wire-loop-based technique for the correction of motion and BCG artifacts, and this method was adopted in real-time (18). (15, 16) suggested adding reference electrodes attached to a conductive reference layer to record artifacts and further remove them from EEG data (see also Luo et al., 2014). Although these methods appear beneficial for reducing artifacts, they are not yet widely used due to their required hardware modifications and additional equipment (21). Unfortunately, these approaches cannot be applied to existing datasets that were recorded without the extra sensors. Third, another artifact reduction strategy uses blind source separation (BSS) for decomposing the EEG data into independent components (ICs, e.g. (22)) and reconstructing the EEG data after removing artifactual ICs (6, 23-30). In addition to those 3 main strategies, other methods such as using deep learning (31), Wavelet transform (32), dictionary learning (33), advanced filters such as Kalman (34), adaptive OBS (35), despiking technique (36), and Bayesian filtering (37) are proposed for reducing MRI-related artifacts. However, the accuracy of those methods has not been validated beyond the groups they initially proposed them (38).

While AAS/OBS and using extra sensors have proven successful for reducing MRI-related artifacts, these methods do not remove ocular and muscle artifacts. Also, BSS approaches are not recommended as the sole approach for reducing such artifacts and they are often combined with OBS or AAS to remove residual gradient and BCG artifacts (23, 39, 40). More specifically, using BSS as the primary method for reducing BCG artifacts is not recommended due to the difficulty in distinguishing BCG components from event-related ones and neural activities (40, 41).

In this study, we proposed an automated pipeline for EEG artifact reduction during fMRI (APPEAR). The APPEAR comprehensive approach is an OBS/AAS-ICA-based algorithm for reducing BCG and gradient artifacts, in addition to motion, ocular and muscle artifacts, designed for 1) substantially improving EEG data quality acquired during fMRI; and 2) making it possible for automated, non-human biased, and faster than manual EEG pre-processing of large EEG-fMRI datasets composed of hundreds of subjects (e.g., Tulsa 1000 (42) and CoBRE studies (43)). APPEAR makes extensive use of EEG-fMRI signal processing functions implemented in the EEGLAB open-source toolbox (44).

## 2. Methods

### *2.1. APPEAR*

The APPEAR algorithm combines OBS/AAS, filtering, and ICA to reduce common types of artifacts contaminating EEG data recorded simultaneously with fMRI.

#### *2.1.1 AAS/OBS and Filtering*

Figure 1A shows the algorithm's first step and procedure for reducing noise and artifacts from EEG data. APPEAR first pre-processed raw simultaneous EEG-fMRI data by removing the gradient artifact, using the OBS included in EEGLAB's FMRIB plugin and function fmrib_fastr (2, 9, 44). The raw EEG data included the slice trigger markers (e.g., R128). Prior to running OBS, volume start was added by setting markers at every n-th occurrence of the slice trigger, where n was equal to the number of slices per volume. Volume trigger timing was used to generate an artifact template in OBS. After the removal of the gradient artifact, the data were downsampled to a 250 S/s sampling rate (4 ms interval, the initial sampling rate of the data was 5,000 S/s). The EEG data were bandpass filtered between 1 and 70 Hz (0.1 and 70 Hz for task-based EEG data) using the built-in FIR filter in EEGLABnamed eegfilt. The fMRI slice selection frequency (19.5 Hz for this study) and its harmonics, vibration noise (26 Hz), and AC power line noise (60 Hz) were removed by band rejection filtering (1 Hz bandwidth).

The AAS algorithm requires identification of the cardiac periods in order to form the artifact subtraction templates (45). To do so, the heartbeat was detected using General Electric (GE) MR-compatible physiological pulse oximetry (with 50 Hz sampling rate), which is a signal collected via a photoplethysmograph with an infrared emitter placed under the pad of the subject's non-dominant index finger. The signal from this device is not sensitive to contamination from MRI environment artifacts, so the heart rate could be accurately detected using the peak detection. In our case, the pulse oximetry signal offered precise detection (by visual inspection) for the heartbeat. Thus, the subsequent analyses use pulse oximetry in the correction process. In our open-access software, we provided two other approaches for selecting heartbeat for the studies which do not collect physiological pulse oximetry data: 1) the FMRIB plugin available with EEGLAB for heartbeat detection using simultaneously-recorded ECG data via the back electrode (2); 2) an automatic cardiac cycle determination approach using ICA (45). This method is partially useful when a signal from the ECG electrode is highly contaminated by MRI environment artifacts or significant movement of subjects. Thus, the identification of cardiac periods could be impractical or difficult to determine.

After detecting the cardiac cycle, BCG artifacts were reduced using AAS, which is included in EEGLAB's FMRIB plugin. Although OBS was reported to outperform AAS for removing BCG artifacts in several studies (46, 47), it could potentially remove some neural activity, as it is shown in Supplementary Figure S1, for data from two different participants. Therefore, we selected AAS as the template correction approach for BCG correction. Additionally, the data were then examined for intervals exhibiting significant motion or instrumental artifacts ("bad intervals") using EEGLAB's function, named pop_rejcont, and bad intervals were marked automatically to be removed for ICA decomposition. For detecting bad intervals, the frequency range and threshold value were set to 0.5 and 7 Hz and 8 dB, using pop_rejcont function, respectively.

*2.1.2 ICA*

After the pre-processing and the removal of the gradient and BCG artifacts (Figure 1A), the following steps (illustrated in Figure 1B) were applied for automatic artifact reduction using ICA. The Infomax ICA algorithm (48), implemented in the EEGLAB toolbox, was applied to the EEG data after the template artifact correction. The ICA algorithm was used to decompose the $N \times M$ EEG data into $L \times M$ ICs, where $N$, $L$, and $M$ denote, respectively, the number of channels, ICs to be estimated, and time-samples. The number of components was set to the number EEG channels (31 for this study). The relationship between the EEG data, **x**, and the ICs, **S**, is given by equation [1]:

$$[1] \qquad \boldsymbol{x}_{[N\ x\ M]} = \boldsymbol{A}_{[N\ x\ N]} \cdot \boldsymbol{S}_{[N\ x\ M]}$$

where **A** is the mixing matrix that carries the coefficients of the linear combination between the EEG data and the ICs (49). Bad intervals could significantly affect the ICA algorithm ability to isolate typical artifacts such as eye blinks (50). Therefore, they were removed prior to ICA, resulting in a new $N \times K$ matrix, **x'**. An ICA was applied, forming a new relationship between the shortened EEG data and the resulting ICs, **S'**, given by equation [2]:

$$[2] \qquad \boldsymbol{x'}_{[N\ x\ K]} = \boldsymbol{A}_{[N\ x\ N]} \cdot \boldsymbol{S'}_{[N\ x\ K]}$$

Creating sharp fluctuations due to excluding bad intervals would not affect ICA performance since ICA algorithms use the spatial information only and many of them shuffle the time point for getting the best results (51). We provided an example of running ICA on shuffled and original-ordered EEG data and compared the results in supplementary materials (Figure S2).

*2.1.2.1 Automatic IC Classification*

ICs were flagged within the APPEAR algorithm if they were determined to be one of the following artifacts: BCG, blink, saccade, single-channel, or muscle. Artifacts are determined with spectrum properties, topographic map properties, or an analysis of each IC's contribution (Wong et al., 2016).

**BCG IC Identification**

BCG artifacts obscure EEG signals recorded inside the MRI scanner, independently of MRI acquisition presence, and significantly affect the EEG data quality. These artifacts occur because the movement of electrically conductive material in a static magnetic field results in electromagnetic induction, as described by Faraday's law. Specifically, motion related to cardiac activity induces electromotive forces in the circuit formed by the EEG recording leads and the scalp contaminating the EEG data with BCG artifacts (5). ICs are flagged as BCG if they meet requirements for the mean power spectral density, topographic map, and IC contribution, as stated in (Wong et al., 2016; Wong et al., 2018). The detailed parameters for identifying BCG components are presented in the Supplementary material.

We modified the protocol for marking the BCG components for removal reported in (Wong et al., 2016; Wong et al., 2018) so that no components showing strong alpha activity in the occipital electrodes were removed. To do so, we defined a template that covered the occipital electrodes (O1, O2, and Oz). If the topographic map had an area overlap (more than 0.4 if unipolar, or 0.91 if bipolar) and if the highest value of the power spectral density (PSD) was in the alpha band range (i.e., 7 to 13 Hz) or if there was an average PSD in the alpha band that was higher than the delta, theta, and beta bands, we did not consider that component to be a BCG artifact. On the other hand, if the topographic map exhibited bipolar properties affecting the right and left hemisphere and having neither the maximum PSD in the alpha band nor the highest average PSD in the alpha band compared to the other EEG frequency bands, we considered that component to be a BCG artifact. Supplementary Figure S3 shows an example of a BCG artifact's IC time series and its features.

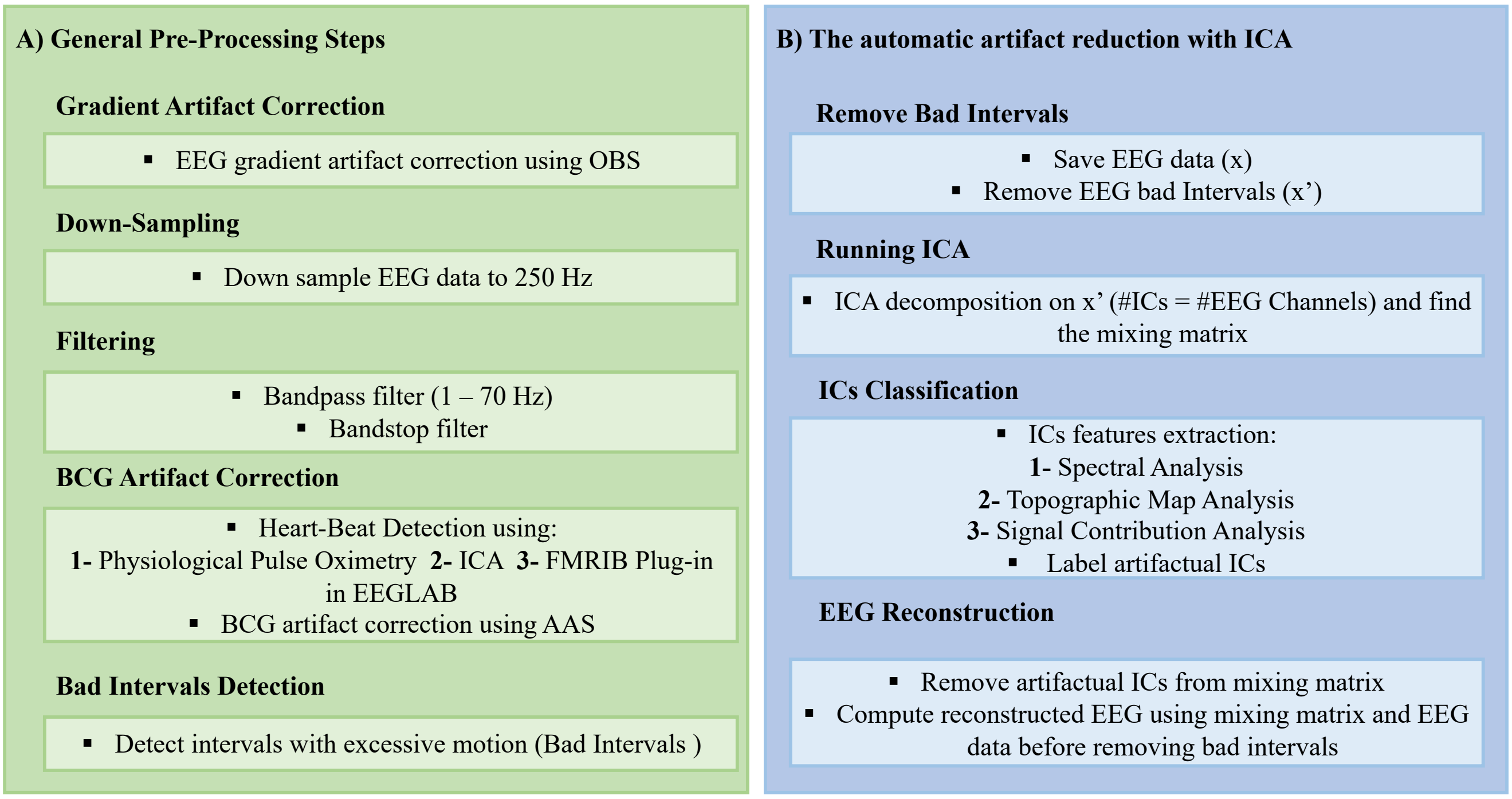

**Figure 1:** The APPEAR Flowchart. Removing EEG artifacts included two main steps: **A)** reducing MRI environment artifacts and filtering. APPEAR first pre-processed raw simultaneous EEG-fMRI data by removing the gradient artifact using the OBS included in EEGLAB's FMRIB plugin and function fmrib_fastr by converting slice trigger markers (e.g., R128) to volume trigger timing and generate a template for gradient artifact. After removing the gradient artifact, the data were down sampled to 250 S/s sampling rate (4 ms interval), and the EEG data were bandpass filtered between 1 and 70 Hz using the built-in FIR filter in MATLAB named eegfilt. The fMRI slice selection frequency, and its harmonics, vibration noise (26 Hz), and AC power line noise (60 Hz) were removed by band rejection filtering (1 Hz bandwidth). In order to find the cardiac cycle for generating BCG artifact template, three methods were offered: 1) physiological pulse oximetry, the signal from this device is not sensitive to contamination from MRI environment artifacts, so the heart rate could be accurately detected using the peak detection. 2) FMRIB plugin for EEGLAB implemented in MATLAB for heartbeat Detection using simultaneously recorded ECG data via the back electrode; 3) Automatic cardiac cycle determination approach using ICA. After detecting the heartbeat events, the BCG artifacts were reduced using AAS, included in EEGLAB's FMRIB plugin. Next, the data were examined for intervals exhibiting significant motion or instrumental artifacts ("bad intervals") using EEGLAB's function, named pop_rejcont, and bad intervals were marked to be further removed. **B)** Independent Component Analysis (ICA). The Infomax ICA algorithm, implemented in the EEGLAB toolbox, was applied to the EEG data after template artifact correction. The ICA algorithm was used to decompose the EEG data into the independent components (ICs). The number of components was set to the number of EEG channels (31 for this study). The bad intervals may have significantly affected the ICA results due to their high amplitude and power. Therefore, they were removed prior to ICA. ICs are flagged within the APPEAR algorithm if they were determined to be one of the following artifacts: BCG, blink, saccade, single-channel, or muscle. Artifacts are determined with spectrum properties, topographic map properties, or an analysis of each IC's contribution. Using the mixing matrix after the bad interval removal and the EEG data before the bad interval removal, an IC matrix related to the whole dataset (before removing bad intervals) was computed. The columns related to artifactual ICs were removed from the mixing matrix and replaced with zero vectors to form a new mixing matrix. Then, a final, reconstructed EEG data matrix, xfinal, with the same size as the original raw EEG, was computed using the original ICA relationship for EEG data and ICs.

### Blink and Saccadic IC Identification

Ocular artifacts are classified as either a blink or a horizontal saccade component. The ICs associated with blinks, as well as saccade, have unique topographic maps. For detecting ICs with topographic maps related to blink and saccade, we used the approach presented in (52). Blink ICs can be identified by their strong spatial projection in the frontopolar area (electrodes Fp1, Fp2). A topographic map related to horizontal saccade ICs exhibits two strong and opposite polarity spatial projections in the orbitofrontal areas (electrodes F7, F8). The details of identifying the topographic maps associated with these two artifacts are presented in the supplementary material of (52). Supplementary Figure S4 shows an example of a blink artifact's IC time series and its features.

### Single-Channel IC Identification

A large artifact can be generated in one channel without affecting any other channels if that channel has a poor electrical connection to the scalp (53). This may result in i) large random low-frequency signal variations, ii) sharp steps in the waveform (electrode pop), and iii) excessive residual MRI gradient artifact for that channel. The component marked as single-channel IC if the Kurtosis of the component is higher than 4, the projection of that component's spectrum power of one channel is significantly higher than the other channels (>5 times), and the power spectrum peak wouldn't be in the alpha band.

### Muscle IC Identification

Muscle electrical activity or "electromyogenic" (EMG) artifacts exhibit widespread high-frequency activity due to asynchronous motor action units (7, 53). These components are flagged if the signal's power is spread out in frequencies higher than 30 Hz, known as the gamma band. Specifically, the average power of the gamma band is computed for each IC, and if the average power is largest in the 30-60 Hz range, the IC is labelled as a muscle artifact (see Supplementary Figure S5 for an example of a muscle artifact's IC features). Such classification considers possible components with a large peak in the gamma band, which typically represent some type of noise (e.g., vibration noise and line noise).

*2.1.2.1 Reconstructing EEG Data after ICA Decomposition*

Using the mixing matrix after bad interval removal (i.e., "$\boldsymbol{A}$") and the EEG data before bad interval removal (i.e., "$x$"), the IC matrix related to the whole dataset (before removing bad intervals) was computed with the following matrix multiplication, given by equation [4]:

[4] $$\boldsymbol{S}_{[N\ x\ M]} = \boldsymbol{A}^{-1}{}_{[N\ x\ N]} \cdot \boldsymbol{x}_{[N\ x\ M]}$$

The columns related to artifactual ICs were removed from the mixing matrix, $\boldsymbol{A}$, and replaced with zero vectors to form a new mixing matrix, $\boldsymbol{A'}$. Then, a final reconstructed EEG data matrix, $\boldsymbol{x_{final}}$, (with the same size as the original raw EEG data) was computed using the original ICA relationship between the EEG data and ICs, given by equation [5]:

[5] $$\boldsymbol{x_{final}}{}_{[N\ x\ M]} = \boldsymbol{A'}_{[N\ x\ N]} \cdot \boldsymbol{S}_{[N\ x\ M]}$$

*2.2 Data Acquisition*

The data used for validation was selected from the Tulsa 1000 (T-1000) study, which assessed and longitudinally tracked 1000 adults, including healthy comparisons and treatment-seeking individuals with mood and anxiety disorders (42). We selected 47 healthy control participants (24 females and 23 males, ranging from 18 to 54 years) of that study. he study was conducted at the Laureate Institute for Brain Research with a research protocol approved by the Western Institutional Review Board (IRB). All volunteers provided written informed consent and received financial compensation for their time to participate in this study.

A GE Discovery MR750 whole-body 3T MRI scanner (GE Healthcare, Waukesha, Wisconsin, USA) and a standard 8-channel, receive-only head coil array were used for fMRI imaging. A single-shot gradient-recalled echoplanar imaging (EPI) sequence with Sensitivity Encoding (SENSE) (Pruessmann et al., 1999) was used for fMRI acquisition (parameters: FOV/slice thickness/slice gap = 240/2.9/0.5 mm, 39 axial slices per volume, 128 × 128 acquisition matrix, repetition time (*TR*), echo time (*TE*) *TR*/*TE*= 2000/27 ms, acceleration factor $R = 2$, flip angle = 90°, sampling bandwidth = 250 kHz). EEG signals were recorded simultaneously with fMRI using a 32-channel MR-compatible EEG system from Brain Products GmbH. An MR-compatible EEG cap (BrainCap-MR) included 32 channels, arranged according to the international 10-20 system. One electrode was placed on the subject's back to record the electrocardiogram (ECG) signal. A Brain Products SyncBox device was used to synchronize the EEG system clock with the 10 MHz MRI scanner clock. The EEG acquisition's temporal resolution was 0.2 ms (i.e., 16-bit 5 kS/s sampling) and measurement resolution for EEG data was 0.1 μV. EEG signals were hardware-filtered throughout the acquisition in the frequency band between 0.016 Hz and 250 Hz.

*2.3 Evaluation*

APPEAR was validated using both an event-related potential (ERP) and resting-state EEG datasets recorded simultaneously with fMRI. We used manually corrected EEG data as a comparison for evaluating the accuracy of the proposed automated pipeline for removing artifacts. We followed the method using template subtraction, followed by ICA, which was suggested for removing EEG artifacts in previous works (6, 23, 40, 41, 54). BrainVision Analyzer 2 software (Brain Products GmbH, Germany) was used to remove the artifacts manually in the semi-automatic mode. The results of manual correction were employed as a reference to evaluate the performance of APPEAR. The five steps procedure for offline EEG artifact reduction was as follows (23). First, imaging artifacts were reduced using the AAS method (9), and the signals were down sampled to 250 S/s. In the second step, band-rejection filters (1 Hz bandwidth) were used to remove the fMRI slice selection fundamental frequency (19.5 Hz in this case) and its harmonics, vibration noise (26 Hz in this case), and AC power line noise (60 Hz). The EEG and ECG data were bandpass filtered from 0.1 to 80 Hz and 0.1 to 12 Hz (48 dB/octave), respectively. In the third step, in order to remove the BCG artifact using AAS (10), the cardiac cycle was automatically detected by the Analyzer 2 software with a subsequent visual inspection that corrected incorrectly positioned R-peak markers. A template of BCG artifacts from 21 preceding cardiac periods for each channel was used to remove BCG artifacts using AAS. In the fourth step, prior to running ICA, the data were carefully examined to exclude intervals exhibiting significant motion or instrumental artifacts. Finally, in the fifth step, the Infomax algorithm (48) was used for ICA decomposition. ICs associated with artifacts were identified based on the topographic map, power spectrum, time course, and kurtosis value. After selecting the artifactual ICs and removing them, the EEG signal was reconstructed with inverse ICA.

*2.3.1 Stop-Signal ERP*

The first dataset used to examine the quality of the corrected data was EEG-fMRI data during a stop-signal task (55) lasting 8 minutes and 32 seconds. To determine the success of the pipeline in the separation and removal of BCG artifacts from EEG data, it is recommended that the quality of the signal of interest is examined (56). Thus, examinations of ERP extracted from the EEG data were used to evaluate the efficacy of the automated processing pipeline. Specifically, data resulting from the automated pipeline were compared to the same data processed manually, as described above. For the ERP analysis, a commonly used paradigm (i.e., stop-signal; e.g., (55)) was employed. During this task, participants were asked to respond to an "X" and "O" with either a right or left button press, but on 25% of the trials, an auditory tone (i.e., "stop-signal") indicated they should not respond. In this paradigm, the stop-signal stimulus was shown to elicit the N2 and P3 waveforms (57-59). The N2 component is a negative deflection in the ERP waveform, maximal over the fronto-central portion of the scalp peaking between 200 and 250 ms (e.g. (60)), and is an indicator of attentional control. The P3 is a centro-parietally maximal positive deflection in the ERP waveform peaking between 300 and 500 ms, which indexes attention allocation (see:(61)). In the current study, the eight participants completed the stop-signal paradigm during simultaneous EEG-fMRI data collection. The analysis was focused on the ERP response to the stop-signal (72 trials for each participant).

In addition to the automated processing pipeline, the data were segmented from 200 ms prior to the 800 ms post onset of the stop-signal. Then the data were baseline corrected to the average of the 200 ms interval preceding the stimulus onset. A low-pass filter was applied to the data with a half-amplitude cutoff of 30 µV and 48 dB/octave roll-off. Finally, automated routines were used to detect bad intervals in the data. Bad intervals were defined as any change in amplitude between data points that exceeded 50 µv; absolute fluctuations exceeded 200 µV in any 200 ms interval of the segments (i.e., −200 to 800 ms); and flat-lining was defined as any change of less than 0.5 µV in a 200 ms period. Trials were excluded if they included any of these artifacts. The number of trials rejected due to the above features ranged from 0 to 10 (mean 3.75, *SD* = 3.24).

According to recommendations from (56), we examined the scalp topographies, waveforms, and peak amplitude measures of the resulting ERP waveforms as well as the estimated signal-to-noise ratio (SNR) of the N2 and P3 waveforms. The SNR of the ERP components was estimated in accordance with recommendations for processing EEG/ERP data (62, 63). Specifically, SNR was calculated for two methods of quantifying ERP amplitude; peak amplitude (a measurement of the largest amplitude a waveform achieves in a specified measurement window) and grand average amplitude (average of the ERP waveform in a specified measurement window). This was done to account for common amplitude measures used in the field to compare groups and conditions in ERP experiments. For peak amplitude, the SNR was calculated as the ratio of the ERP component peak and the difference between the largest negative peak and largest positive peak in the pre-stimulus baseline (estimate of noise). The grand average amplitude SNR was calculated as the ratio of the mean amplitude measured across the following time windows, with respect to stimulus onset: N2, 175 to 225 ms; P3, 300 to 500 ms, to the noise estimate in the baseline period (i.e., −200-0ms) described above. All statistical analyses were conducted in R version 3.6.1 using the WRS2 package.

*2.3.2 Resting-State*

A resting-state EEG-fMRI run, lasting 8 minutes, was conducted for each subject. Prior to the rest run, participants were instructed to clear their minds, not think about anything in particular, and try to keep their eyes open and fixated on a cross. In order to evaluate the resting-state EEG data quality using our proposed pipeline, we compared the time-frequency (Wavelet Transform) and spectral power (FFT) results between the manually corrected and automatically corrected EEG data.

The Continuous Wavelet Transformation (CWT) was applied to the data after taking the average EEG signal among all channels (i.e., 31 channels). CWT deployed the analytic Morse wavelet implemented in MATLAB's function cwt, with symmetry parameters of 3 and a time-bandwidth product of 60. To compare the results between the manually- and APPEAR-corrected EEG sets, we plotted the time-frequency analysis for only a 30-second segment of the EEG recording (for a better visibility) taken from 60 to 90 seconds.

In addition, we computed the average power spectral density (PSD) for all EEG channels for both manual- and APPEAR-corrected data. To calculate the PSD in each analysis and each channel, a moving window FFT, with 4.096 s data interval length (0.244 Hz spectral resolution) and 50% interval overlap with a Hanning window, was applied after the exclusion of the motion-affected intervals marked manually during the manual correction. After that, the PSD was averaged among all channels for each subject.

## 3. Results

The APPEAR pre-processing run times for each individual subject are shown in Table 1. The run time was measured in terms of the time to run the entire process on MATLAB 2016B on an Intel Core i5-7500T 2.7GHz workstation with 8 GB RAM (Model: Lenovo ThinkCentre M710q) and Windows 10. When compared to manual correction, which could take up to hours, APPEAR took less than 15min/subject.

The percentage length of the original signal was marked as bad segments on an average across the different sessions for the EEG data are as follows (mean ± standard deviation): Rest: 15.7 ± 8.4 sec; Task: 14.04 ± 10.18 sec.

Comparisons of the resulting ERP components between the APPEAR and manually processed data are presented in Figure 2.

**Table 1.** APPEAR EEG pre-processing computation times (Run Times in seconds) for each subject for rest (8 minutes) and task (Stop-Signal, 8 minutes and 32 seconds) EEG-fMRI datasets.

| | **Run Time (seconds)** | |
|---|---|---|
| | **Rest** | **Task** |
| **Subject 1** | 616 | 712 |
| **Subject 2** | 628 | 702 |
| **Subject 3** | 650 | 679 |
| **Subject 4** | 659 | 654 |
| **Subject 5** | 688 | 735 |
| **Subject 6** | 688 | 723 |
| **Subject 7** | 700 | 697 |
| **Subject 8** | 742 | 741 |

Table 2 includes the means (*M*), standard deviation (*SD*), and statistical comparison (i.e., dependent samples *t*-test) of the means of the mean amplitude ERP components (i.e., N2, P3) between automated and manually corrected ERP data. Results indicate that there are no significant differences between mean amplitude ERPs calculated from data resulting from the automated pre-processing (i.e., APPEAR) and those calculated following manual pre-processing (uncorrected *p*-values range from 0.07 to 0.40 and Cohen's effect size *d* range from 0.06 and 0.23). It should be noted that the N2 was quantified as the mean amplitude in midline channels (i.e., Fz, FCz, Cz) between 175 and 225 ms post-stimulus onset, based on a combination of visual inspections of the current data and previous research, indicating the N2 peaks between 200-250ms (60). Notably, the N2 peak was not evident at Pz. The P3 was calculated as the mean amplitude between 300 and 500 ms post-stimulus onset at midline channels (i.e., Fz, FCz, Cz, Pz). Figure 2 represents the ERP component waveforms and scalp topographies for both manually and automated corrected data.

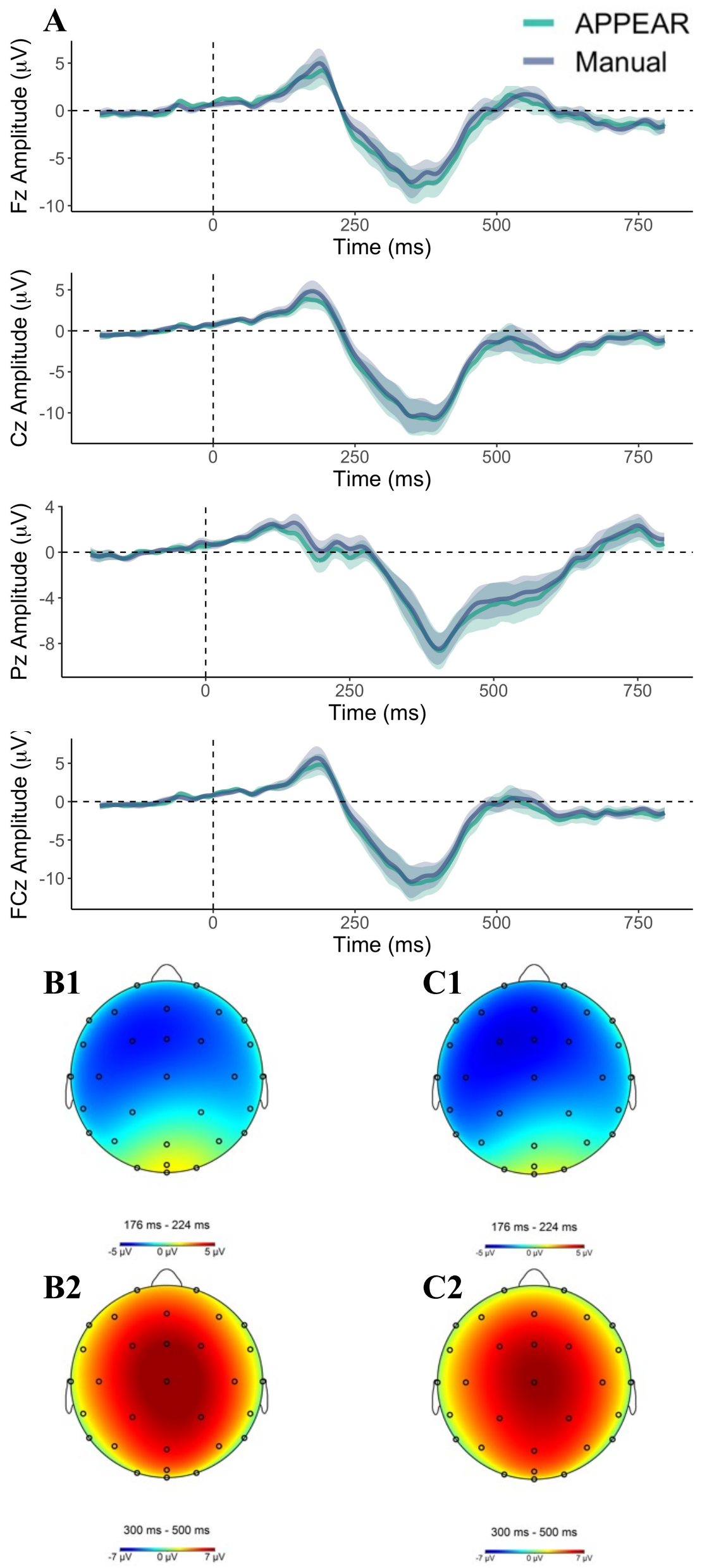

**Figure 2:** Averaged waveforms and topographical maps for ERP waveforms (i.e., N2, P3) using APPEAR and Manual corrections among 8 participants. **A)** ERP Waveforms comparing automated (green) and manual (blue) pre-processing pipelines are displayed at all midline measurement electrodes. Time-zero represents the onset of the auditory stop-signal stimulus. Shaded areas represent the standard error of the mean for the ERP signal at each time point. Presented waveforms were calculated from average mastoid referenced EEG data. **B1)** N2 scalp topography from the automated pipeline represents average activation across the scalp during the measurement window relative to the 200 ms pre-stimulus baseline. **B2)** P3 scalp topography from the automated pipeline represents average activation across the scalp during the measurement window relative to the 200 ms pre-stimulus baseline. **C1)** N2 scalp topography from the manual pipeline represents average activation across the scalp during the measurement window relative to the 200 ms pre-stimulus baseline **C2)** P3 scalp topography from the manual pipeline represents average activation across the scalp during the measurement window relative to the 200 ms pre-stimulus baseline.

**Table 2.** T-tests comparing mean amplitude N2, P3 across automated (APPEAR), and manual (Manual) processing. Cohen's d value was calculated for each comparison. The mean, M, and standard deviation, SD, measurements are in µV.

| N2 | APPEAR M(SD) | Manual M(SD) | Mean comparison |
|---|---|---|---|
| Fz | 1.09(1.09) | 1.33(1.37) | $t(7) = 1.03, p = 0.34, d = 0.19$ |
| FCz | 1.32(1.25) | 1.54(1.66) | $t(7) = 1.01, p = 0.35, d = 0.14$ |
| Cz | 1.50(1.87) | 1.30(1.46) | $t(7) = -0.96, p = 0.37, d = 0.12$ |
| **P3** | | | |
| Fz | 1.61(1.14) | 1.59(1.44) | $t(7) = 0.08, p = 0.93, d = 0.02$ |
| FCz | 2.21(1.56) | 2.35(2.20) | $t(7) = -0.35, p = 0.73, d = 0.07$ |
| Cz | 3.56(4.28) | 2.65(2.28) | $t(7) = 1.31, p = 0.23, d = 0.26$ |
| Pz | 2.37(2.23) | 2.17(1.71) | $t(7) = 0.84, p = 0.43, d = 0.001$ |

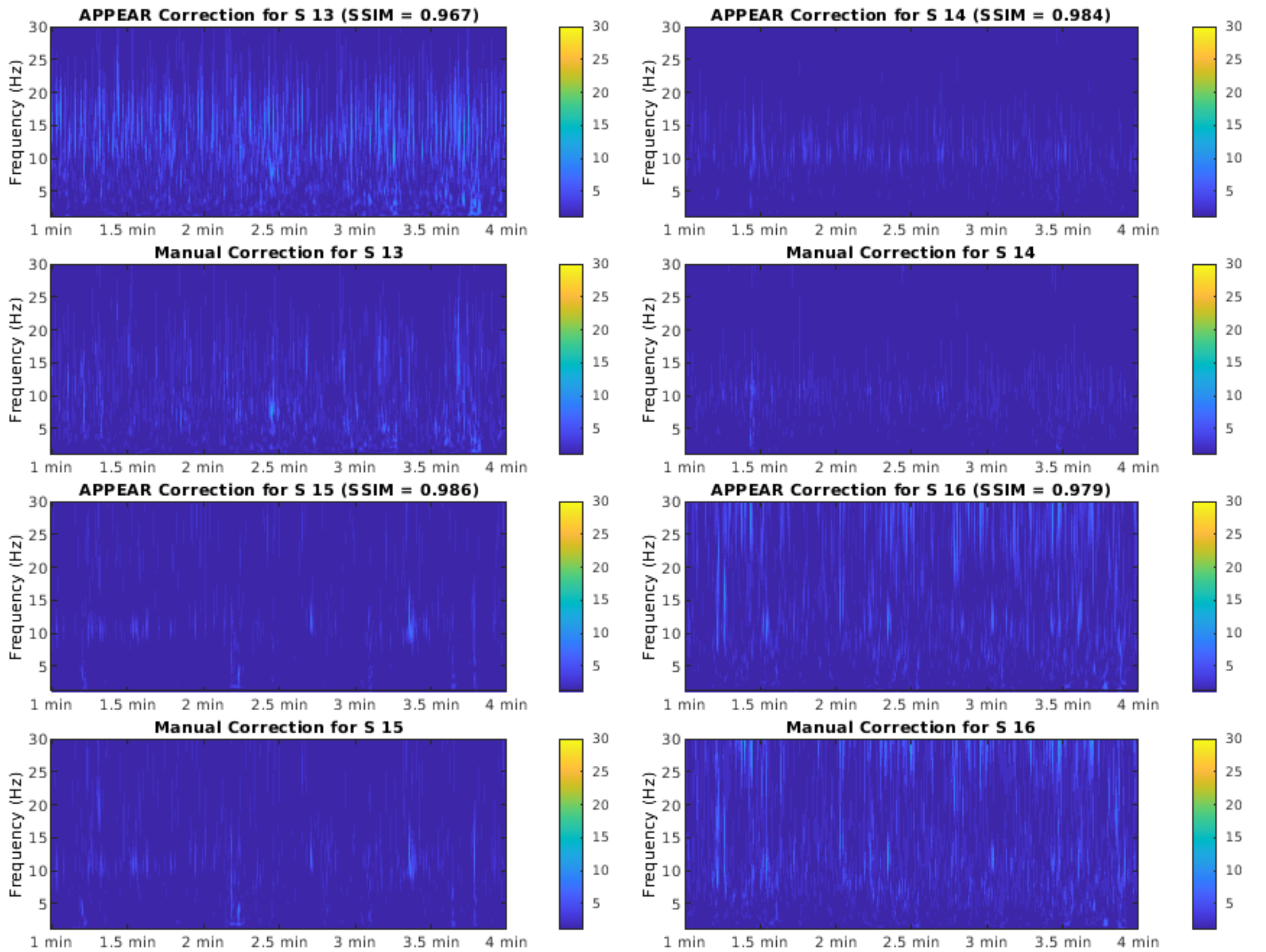

**Figure 3:** The time/frequency comparison (Wavelet) between APPEAR and Manually Corrected EEG Data for 4 Subjects (the plots for the rest of subjects are provided in supplementary figures S5-16). The Continuous Wavelet Transform (CWT) was applied to the data after taking the average EEG signal among all channels (i.e., 31 channels). To compare the results between the manually and automatically corrected EEG sets, we plotted the time-frequency analysis for only a 3-mintute segment of the EEG recording taken from the middle of the EEG recording (60 seconds towards the end of the recording) for each individual subject. The figures for all subjects show a similar pattern for the manually and automatically corrected EEG sets. We used Structural similarity (SSIM) index to compute the similarities between APPEAR and manually corrected images.

**Table 3.** T-tests comparing signal-to-noise ratios N2 and P3 across automated (APPEAR) and manual (Manual) processing. Cohen's d value was calculated for each comparison. M and SD represent the mean and standard deviation, respectively.

| | APPEAR M (SD) | Manual M (SD) | Mean comparison |
|---|---|---|---|
| **N2** | | | |
| Fz | -3.31(3.21) | -3.74(3.34) | $t(7) = 2.12$, $p = 0.07$, $d = 0.13$ |
| FCz | -3.89(2.93) | -4.28(3.34) | $t(7) = 1.76$, $p = 0.12$, $d = 0.12$ |
| Cz | -2.74 (2.47) | -3.37(3.00) | $t(7) = 2.03$, $p = 0.08$, $d = 0.23$ |
| **P3** | | | |
| Fz | 5.10(3.01) | 4.44(2.53) | $t(7) = 1.57$, $p = 0.16$, $d = 0.23$ |
| FCz | 7.04(3.76) | 6.50(3.32) | $t(7) = 1.48$, $p = 0.18$, $d = 0.15$ |
| Cz | 7.62(4.05) | 7.21(3.55) | $t(7) = 1.10$, $p = 0.31$, $d = 0.11$ |
| Pz | 5.51(4.11) | 5.26(3.52) | $t(7) = 0.90$, $p = 0.40$, $d = 0.06$ |

Table 3 includes the means, standard deviation, and statistical comparison (i.e., dependent samples *t*-test) of the SNRs of the mean amplitude ERP components (i.e., N2, P3). A series of dependent sample *t*-tests presented in Table 3 indicates that there were no significant differences between ERP components (i.e., N2, P3) resulting from the automatic processing compared to the manual processing (uncorrected *p*-values range from 0.23 to 0.93 and Cohen's effect size range from 0.02 and 0.26).

Table 4 represents the estimated SNRs of the mean amplitude and peak amplitude measurements from the grand average waveform across subjects of the N2 and P3 waveforms (i.e., N2, 175 to 225 ms; P3, 300 to 500 ms, post-stimulus onset).

**Table 4.** Signal-to-noise ratios in the grand average waveforms (among 8 participants) obtained with automated (APPEAR) and manual (Manual) processing.

| **APPEAR** | **Fz** | **Cz** | **Pz** | **FCz** |
|---|---|---|---|---|
| GA Peak N2 | 2.866 | 2.604 | - | 3.052 |
| GA Peak P3 | 5.399 | 7.519 | 8.135 | 6.778 |
| GA Mean N2 | 2.239 | 1.899 | - | 2.456 |
| GA Mean P3 | 3.443 | 5.281 | 5.19 | 4.45 |
| **Manual** | | | | |
| GA Peak N2 | 5.128 | 3.556 | - | 4.046 |
| GA Peak P3 | 7.741 | 7.787 | 5.794 | 7.442 |
| GA Mean N2 | 3.597 | 2.308 | - | 2.853 |
| GA Mean P3 | 4.577 | 5.296 | 3.595 | 4.628 |

Figure 3 shows the comparison between CWT results from APPEAR and manually corrected data for the first 4 participants, and the rest were presented in the supplementary Figures S6-17.

We also compared the PSD (averaged among all channels) in different frequency bands between the APPEAR and manually-corrected EEG using a paired t-test (Figure 4) and the results are as follows: Delta band: $t(46) = -0.02$, $p = 0.99$; Theta Band: $t(46) = 1.34$, $p = 0.19$; Alpha band: $t(46) = 1.84$, $p = 0.07$; Beta Band: $t(46) = 1.30$, $p = 0.20$.

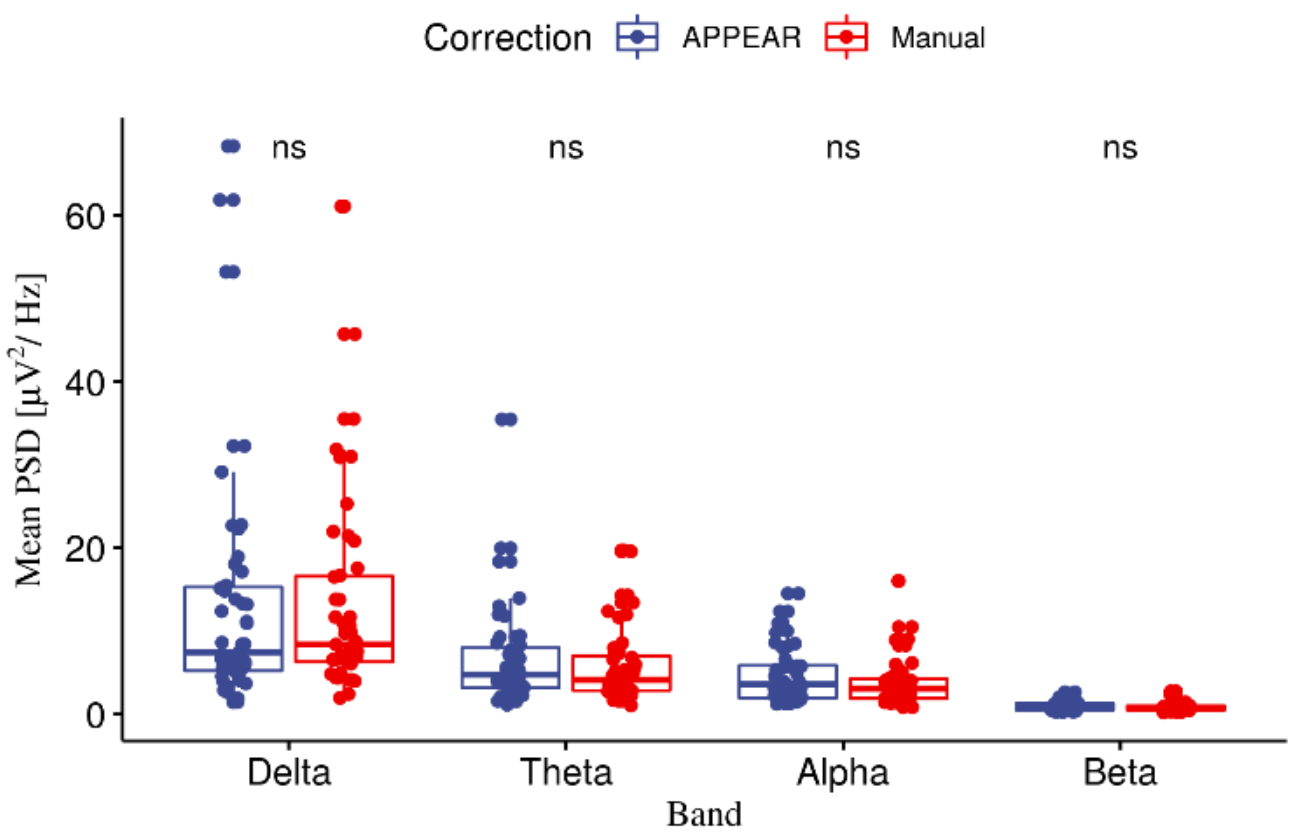

**Figure 4:** The Mean Powers Spectral Density (PSD) Comparison between APPEAR and Manually Corrected EEG Data in delta; theta; alpha; and beta bands. The PSD for all EEG channels for both the manual and APPEAR corrected data were computed in the different EEG frequency band (i.e., delta, theta, alpha, and beta). For calculating the PSD in each analysis and each channel, a Hanning moving window FFT of 1024 sample and 50% interval overlap, was applied after exclusion of the motion-affected intervals marked manually during the manual correction. After that, the PSD was averaged among all channels for each subject in each EEG frequency band.

Furthermore, the association between heart rate and the and average PSD was investigated to examine any influence of heartbeats on the EEG signal (Figure 5).

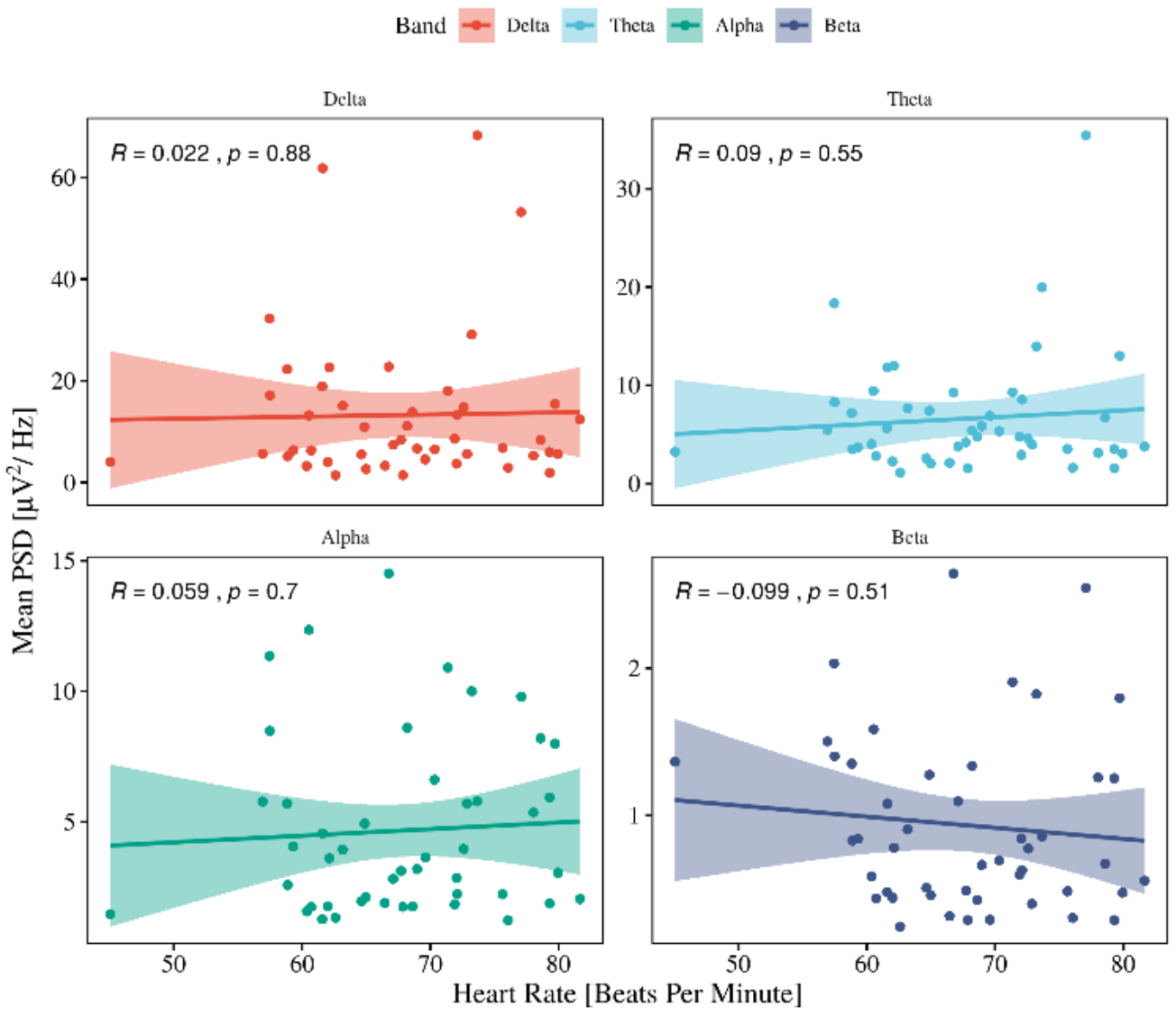

**Figure 5:** The Correlation Analysis between the Heart Rate (HR) and the Mean PSD from Delta Band for APPEAR corrected data. The correlation shows no association between mean PSD for delta band and HR, which indicates that BCG artifacts were detected and suppressed from EEG.

## 4. Discussion

In this work, we proposed a fully automated pipeline for removing EEG artifacts recorded simultaneously with fMRI. The pipeline was validated on both resting-state and task-based datasets by comparing APPEAR-pre-processed and manually pre-processed EEG data.

### MRI Environment Artifact Reduction

Reducing MRI gradients and BCG artifacts is the first step of artifact correction for any EEG data recorded during fMRI acquisition. To do this, we first employed a template artifact correction. In the current study, we noticed a drawback of using OBS instead of AAS. Supplementary Figure S1 illustrates important caveats in using OBS as an average template subtraction method, as it removes some neural activity (e.g., alpha wave in posterior and occipital channels). Therefore, we employed AAS for reducing BCG artifacts instead of OBS.

### Automatic Classification of Artifactual ICs after ICA Decomposition

Classifying the ICs may be the most challenging step in removing EEG artifacts, regardless of being recorded inside or outside of the MRI scanner. Although several methods have been proposed for automatic/semi-automatic IC classification for EEG data recorded outside the MRI (53, 64-71), there are very few for EEG data recorded inside the scanner (23, 72, 73). Here in this study, we classify the components either as artifacts or neural activities. IC classification was determined with spectrum properties, topographic map properties, or an analysis of each IC’s contribution. Using those features, we removed the ICs associated with residual BCG, ocular, muscle, and single-channel artifacts.

### APPEAR Evaluation

In this study, we validated our automated EEG pre-processing pipeline performance for two common applications of simultaneous EEG-fMRI (i.e., resting-state and ERP). For resting-state, we compared the wavelet transformation and FFT results between the manually corrected and APPEAR-corrected EEG data. Our results showed no significant difference between the two approaches. Furthermore, the observed time course and scalp topographies (see: Figure 2) are similar to prior research examining the N2 and P3 in the stop-signal paradigm (e.g., (57-59)) as well as the manually corrected results.

### EEG Pre-processing Speed

Manual pre-processing of the EEG data acquired during fMRI requires both extra time and a trained and experienced researcher, especially when compared to EEG recorded outside of the MRI scanner. The ICA classification might be one of the most challenging steps. The analysis time can vary greatly depending on the characteristics of various neuronal activities in relation to artifacts (e.g., neuronal signal magnitude vs. artifact magnitude). APPEAR offers a comparable quality of EEG pre-processing and artifact suppression, in addition to a much-reduced time requirement per subject. As presented in Table 1, the run time for all APPEAR pre-processing steps is less than 15 minutes per subject (utilizing modest computer hardware as well as software not configured and optimized for computational speed), which is significantly less than the time required for a human researcher to complete the same task. Beyond improved speed, APPEAR makes it possible to and will allow for the pre-processing and suppression of EEG artifacts in clinical EEG-fMRI studies, like the Tulsa 1000 (42), with a large number of participants.

### Limitations and Future Directions

The reported approach has several limitations. First, detecting the cardiac periods is still a challenging part of using template artifact subtraction methods and could influence the efficacy of removing artifacts significantly with either of the aforementioned method. To have the best possible estimate of the cardiac cycles, we used a newly developed technique for detecting the cardiac cycle using ICA on EEG data (Wong et al., 2018). This approach generally outperforms the FMRIB plugin implemented in MATLAB for cardiac cycle detection (Supplementary Table S1). However, we confirmed the estimation of the cardiac cycle using the pulse oximeter waveform (which is unaffected by MRI environment artifacts). If the ICA method did not detect the cardiac periods accurately, then we used the fMRIB approach using the ECG signal recorded via the electrode on the subject's back. The SNR values for APPEAR corrected data presented in Table 4 are lower compared to manually corrected data for N2, although this effect is not significantly different (Table 3).

Second, the implemented procedure for automated classification of independent components makes various assumptions and utilizes a large number of a priori selected numerical parameters (Wong et al., 2016). While these parameters might enable the acceptable performance of the ICA-based artifact correction on average, the performance could be suboptimal for individual cases. The reason is that neuronal activity patterns and various artifacts exhibit large variability across subjects. Careful analyses of the ICA-based artifact correction performance for large cohorts of participants will be required to optimize these parameters and enable their meaningful adjustment depending on an individual's neuronal activity properties and artifact characteristics.

Third, separation of EEG data into independent components is never perfect, which means that multiple ICs typically contain mixtures of neuronal signals and various residual artifacts. Proper classification of such ICs is particularly difficult and necessitates a comprehensive analysis of their properties and contributions to the measured EEG signals. The reported automated IC classification procedure cannot effectively treat such mixed neuronal/artifactual ICs. This problem will require further research and implementation efforts.

In this work, we improved the automatic IC classification compared to the previous real-time EEG artifact correction study (23). However, the computation speed of the algorithm must be further improved to be used in real-time applications.

## 5. Conclusion

The manual removal/suppression of EEG artifacts is one of the main challenges for simultaneous EEG-fMRI experiments because it is both time-consuming and requires specialized expertise. We developed a fully automated pipeline for EEG artifacts reduction (APPEAR). APPEAR was validated and compared to manual EEG pre-processing for two common applications - resting and task-based EEG-fMRI acquisitions. APPEAR correctly removed common EEG artifacts, such as gradient, BCG, eye blinks, motions, and muscle artifacts. APPEAR offers faster pre-processing times as compared to manual processing and provides the capacity and possibility for large-scale EEG pre-processing as well as the analysis of clinical EEG-fMRI datasets composed of hundreds of subjects with affordable time and efforts. In providing a more efficient method of removing EEG artifacts, our work represents an important step and incentive towards expanding EEG-fMRI applications in the study of the human brain both in health and disease.

## Toolbox Availability

Please use the following link to access the toolbox:

https://github.com/obada-alzoubi/appear

**Supplementary Material for:**

# Automated Pipeline for EEG Artifact Reduction (APPEAR) Recorded during fMRI

## Supplementary Figure S1

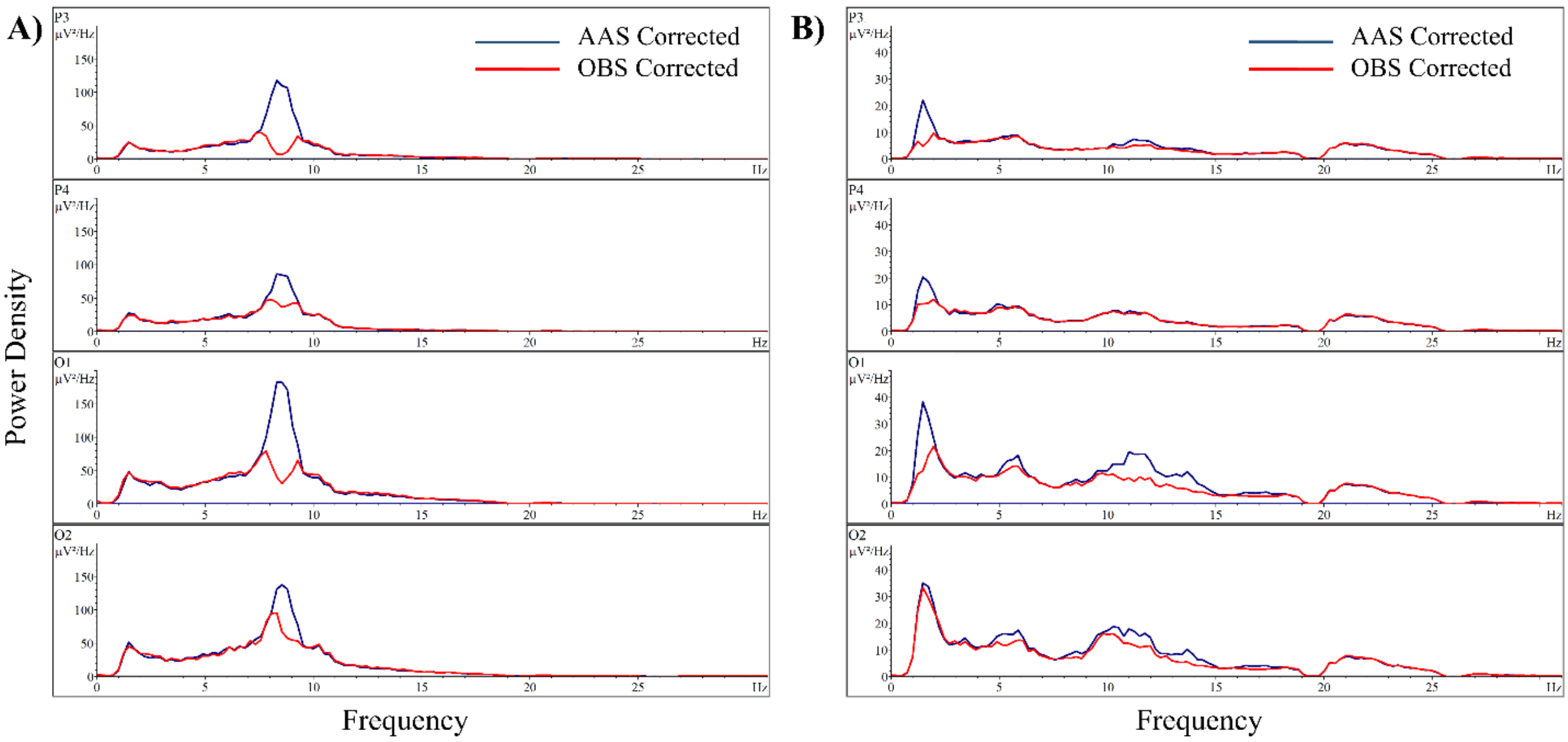

**Supplementary Figure S1:** Comparison between Power Spectral Density (PSD) after applying Average Artifact Subtraction (AAS, black line) and Optimal Basis Sets (OBS, red line) for subject 5 (**A**) and subject 8 (**B**).

**Details for detecting independent components (IC) associated with ballistocardiogram (BCG) artifacts.**

The power spectrum is divided into two ranges, cardioballistic (2-7 Hz) and neuronal (8-12 Hz). If an IC has a cardioballistic artifact, the power spectrum shows peaks in both the cardioballistic and neuronal frequency ranges. The method used by (Wong et al, 2018) determines the rise of the peaks in both regions and requires that they meet four conditions to be considered as a BCG artifact. Condition (i) states that a large peak must be present in the cardioballistic frequency range; Condition (ii) states the Rise of the Neuronal peak ($R_N$) must be small; and if (ii) is not satisfied, then Conditions (iii) and (iv) define comparable spectrum amplitudes required in the cardioballistic and neuronal ranges for a BCG IC. For Condition (ii), to obtain the full $R_N$, a frequency range is defined between the Frequency at the Local Minimum ($f_{LMin}$) immediately below 8 Hz and the Frequency at the Peak ($f_P$) in the neuronal range. If such a local minimum exists, the frequency range is taken as $[f_{LMin}, f_P]$; otherwise the frequency range becomes $[8\text{ Hz}, f_P]$. The power at the lower and upper bound of the frequency range is denoted as $S(f)$ where $f=f_{LMin}$ or $f=8$ Hz and $f=f_P$, respectively. The $R_N$ is calculated as the difference between the $S(f_P)$ and the minimum power given within the frequency range (either $S(f_{LMin})$ or $S(8\text{ Hz})$), calculated with either equation [S1] or [S2] depending on if a local minimum immediately below 8 Hz exists.

**[S1]** $R_N = S(f_P) - \min(S(f), f \in [f_{LMin} f_P])$, if $f_{LMin}$ exists

**[S2]** $R_N = S(f_P) - \min(S(f), f \in [8\text{ Hz}, f_P])$ if $f_{LMin}$ does not exist

For Conditions (iii) and (iv), the minimum power ($S_{min}$) below the neuronal peak frequency is defined as a baseline for each spectrum. A cardioballistic motion IC is recognized when the average power ($S_{ave}$) in the cardioballistic frequency range is comparable to the neuronal peak rise. There may be multiple

Peaks in the CardioBallistic ($P_{cb}$) range, $i$=1,…,$P_{cb}$. Condition (iii) requires that the cardioballistic peaks have a local minimum on the left and a peak rise larger than $0.2S_{ave}$. Condition (iv) requires that the maximum peak Rise in the CardioBallistic range ($R_{cb}$), or the average power over the cardioballistic range ($S_{cb}$) is sufficiently large compared to the Rise of the Neuronal peak ($R_N$). Condition (iv) is met if any of the following equations [S3 – S5] are met:

**[S3]** $R_N <= 0.33\ S_{ave}$

**[S4]** For $P_{cb}$ cardioballistic peaks with local left minimum and $R_{cb,k} > 0.2\ S_{ave}$, where $k$=1,…, $P_{cb}$, $\max(\{R_{cb,k}\}, k=1,\ldots, P_{cb}) > R_N -3$

**[S5]** For $P_{cb}$ cardioballistic peaks with local left minimum and $R_{cb,k} > 0.2\ S_{ave}$, where $k$=1,…, $P_{cb}$, $\mathrm{mean}(S(g), g \in [2\ \mathrm{Hz}, 7\ \mathrm{Hz}]) - S_{min} > 0.33\ R_N$, $\max(\{S_{cb,k}\}, k=1,\ldots, P_{cb}) > S_N -3$

where in [S5] the peak power of the neuronal range is stated as $S_N$.

The spatial projection of each IC onto the EEG channel space forms a topographic map. The spatial projection vector is interpolated using the MATLAB function, griddata. Normally, BCG ICs exhibit bipolar topographies, i.e. opposite polarities for opposite regions (Zotev at al., 2012). During the topographic map analysis, the values are normalized. Then, two sets of polarity regions, primary and secondary, are defined. (Wong et al., 2018) creates polarity arc regions, defined as the overlapping polarity regions using a topographic map boundary with a width of 0.2. Any region not defined by the primary and secondary regions are labeled as neutral regions. Using the three conditions developed in (Wong et al, 2018), the BCG ICs are flagged. Condition (i) requires that there be up to one neutral region in the topographic map; Condition (ii) requires that only one positive (or negative) polarity region and polarity arc region are allowed in the topographic map; Condition (iii) ensures that there is a left/right opposite polarity region with one negative (or positive) primary polarity region and polarity arc region; and Condition (iv) sets the minimum areas for the secondary polarity region and polarity arc region in the topographic map.

In the time-series of a BCG IC, there are distinct peaks (approximately every 1 second) caused by cardiac pulsations. Removing the BCG IC from the EEG time-series signal shows a steady signal reduction at the pulsation peaks. Looking at the signal contribution of a BCG IC, the average positive and negative magnitudes ($\alpha_+$ and $\alpha_-$, respectively) of the reduced signal ($\alpha'$) after removing the IC are compared to the original time-series signal ($\alpha$). In (Wong et al, 2018), the thresholds for the average positive and negative magnitudes for any channel j are: (i) $0.5(\alpha_{j+}'/\alpha_{j+} + \alpha_{j-}'/\alpha_{j-}) < 0.97$ and (ii) $\min(\alpha_{j+}'/\alpha_{j+} + \alpha_{j-}'/\alpha_{j-}) < 0.95$. If these two thresholds are met, the IC is flagged as a BCG artifact.

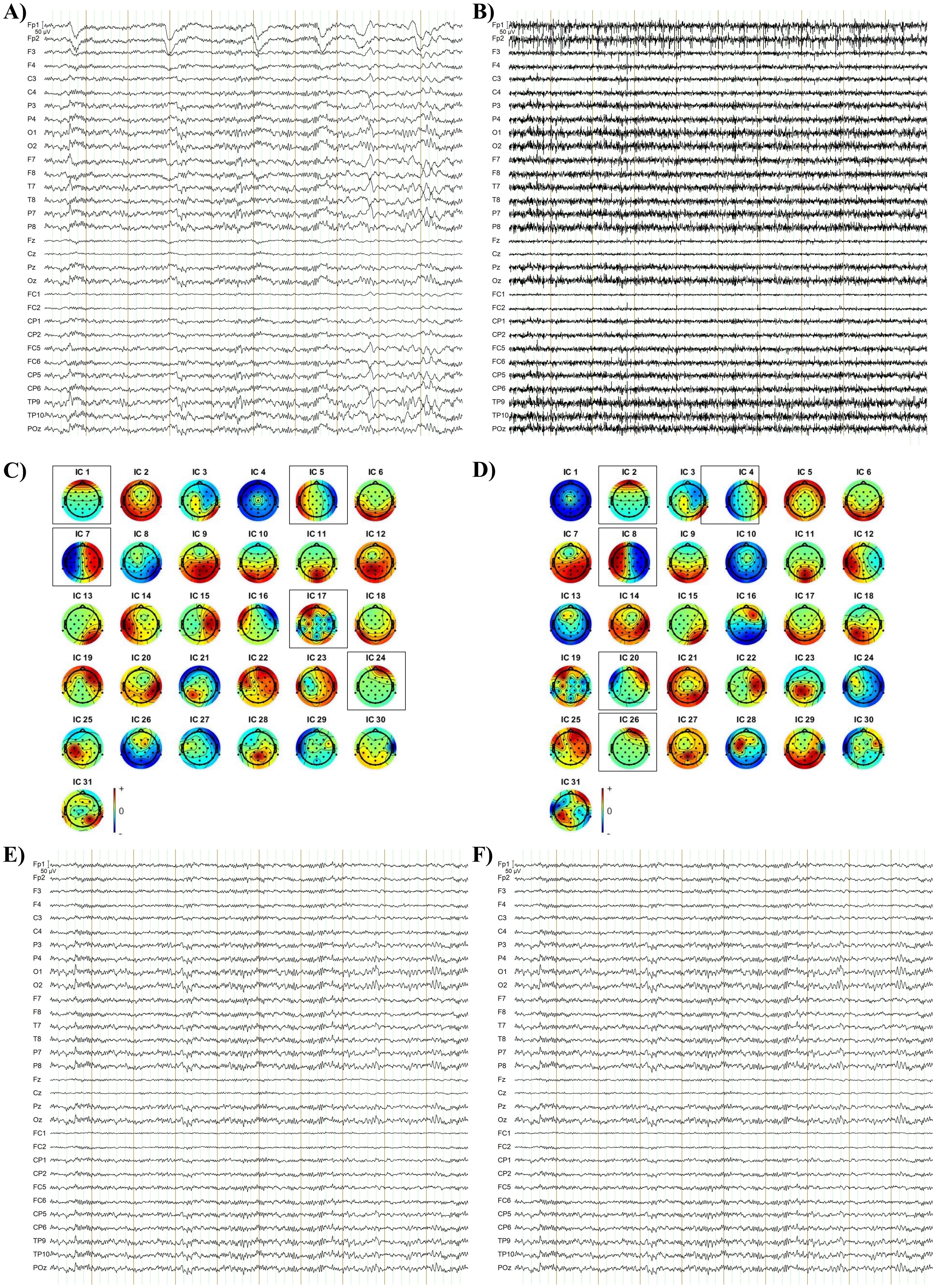
A)
B)
C)
D)
E)
F)
Fp1
50 µV
Fp2
F3
F4
C3
C4
P3
P4
O1
O2
F7
F8
T7
T8
P7
P8
Fz
Cz
Pz
Oz
FC1
FC2
CP1
CP2
FC5
FC6
CP5
CP6
TP9
TP10
POz
IC 1
IC 2
IC 3
IC 4
IC 5
IC 6
IC 7
IC 8
IC 9
IC 10
IC 11
IC 12
IC 13
IC 14
IC 15
IC 16
IC 17
IC 18
IC 19
IC 20
IC 21
IC 22
IC 23
IC 24
IC 25
IC 26
IC 27
IC 28
IC 29
IC 30
IC 31
+
0
-

**Supplementary Figure S2: The figure demonstrates the heuristic property of independent component analysis (ICA).** The time series of 10 seconds EEG data from one participant **(A)** and the data after randomly shuffled (the entire length of EEG data was shuffled, but we just present 10 seconds of the data for better visualization) **(B)** before running independent component analysis (ICA). The topography map of independent components (ICs) after running ICA on the original EEG data **(C)** and shuffled EEG data **(D)**. The artifactual components are marked (they are in black squares). **E)** The EEG data after removing artifactual ICs. **F)** The shuffled EEG data after removing artifactual ICs and reverting to the original sort order (re-ordering was performed after running inverse ICA). Although the order of ICs is different when we applied ICA on the original EEG and shuffled one, in both, we could distinguish the artifactual ICs, and after removing them and reverting the shuffled data to the original sort order, we get the same artifact-reduced results.

## Supplementary Figure S2

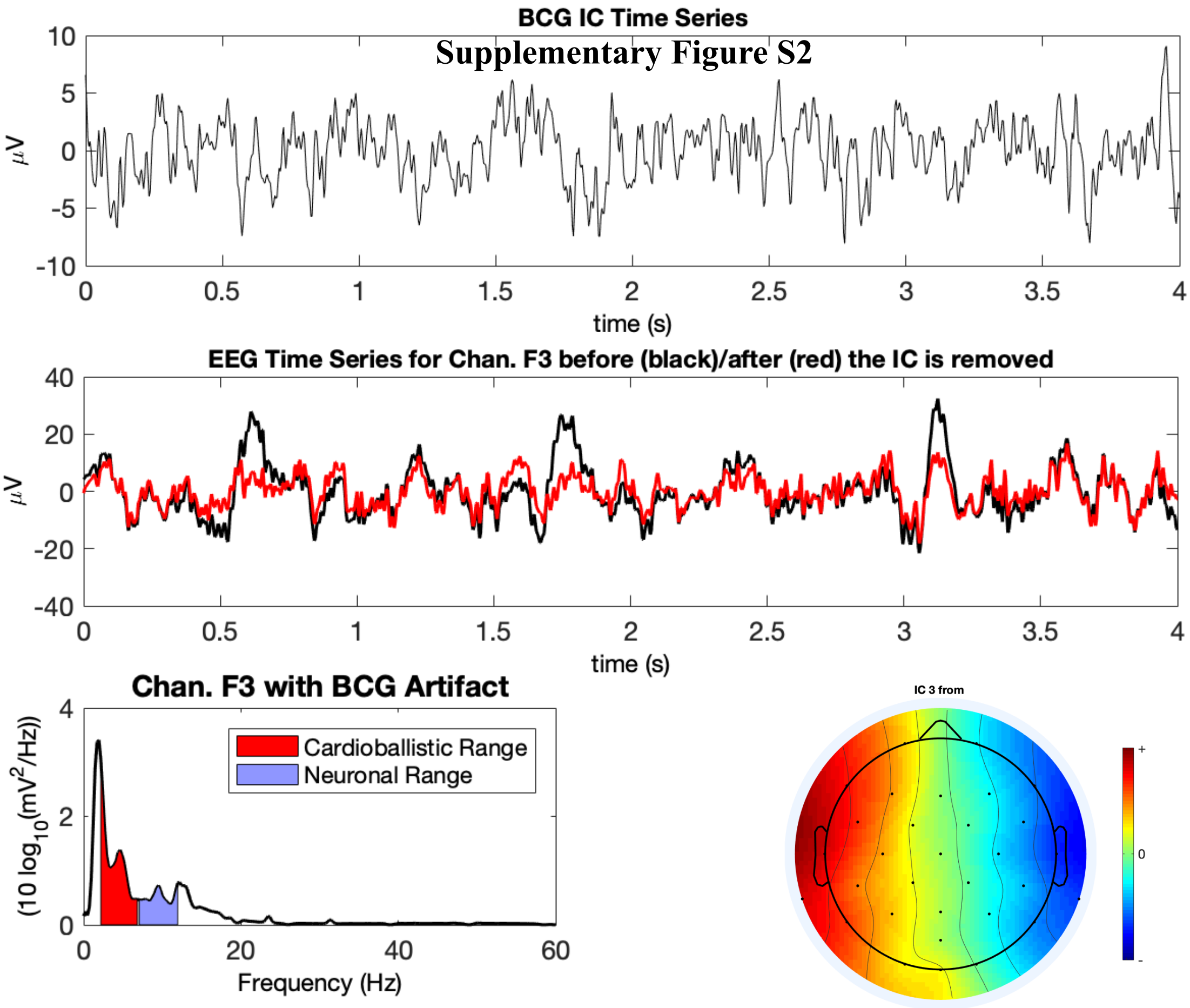

**Supplementary Figure S3:** An example of a Ballistocardiogram (BCG) component.

## Supplementary Figure S3

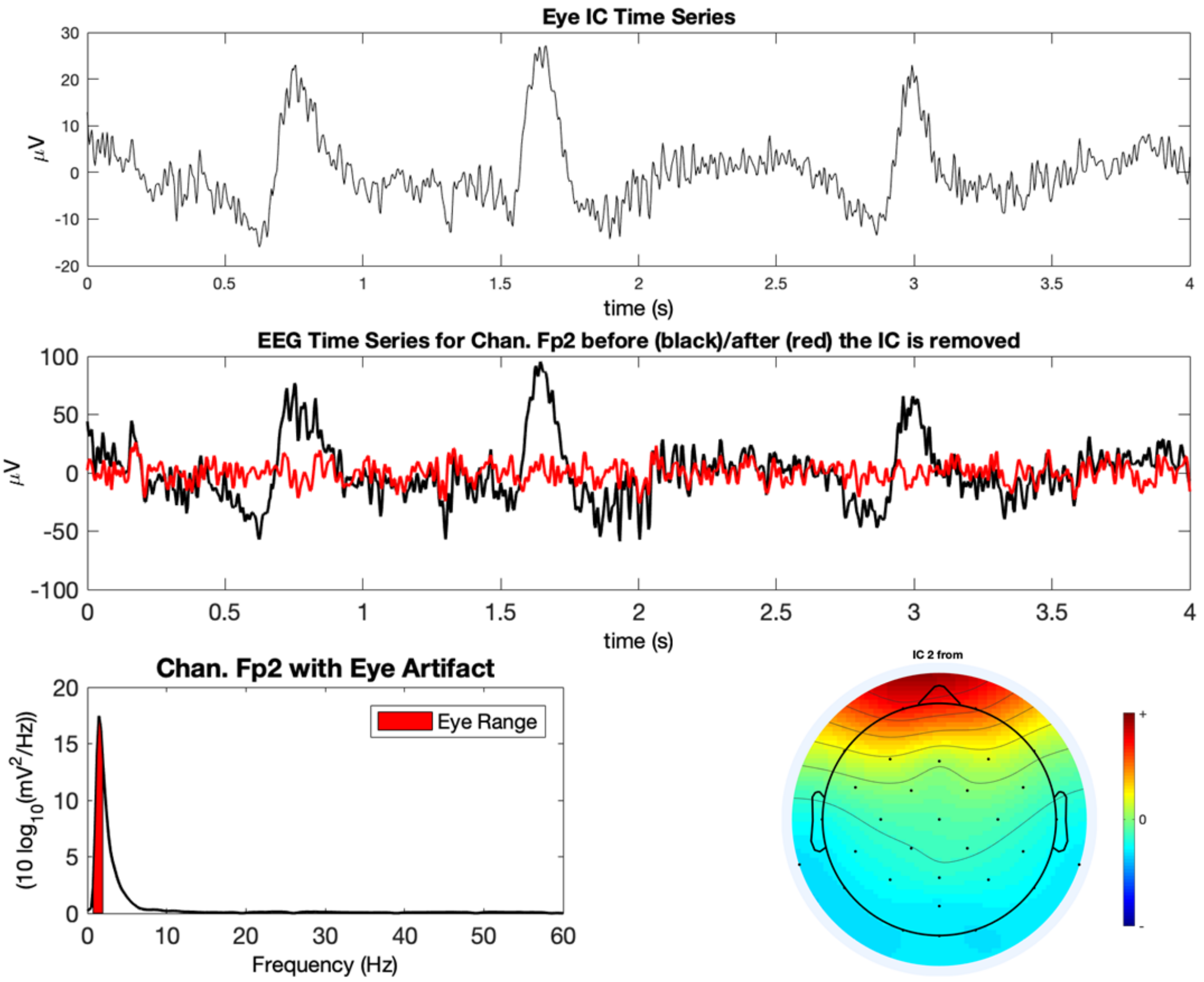

**Supplementary Figure S4:** An example of a Blink Component. Blink ICs can be identified by their strong spatial projection in the frontal area and low frequency activity in delta band.

## Supplementary Figure S4

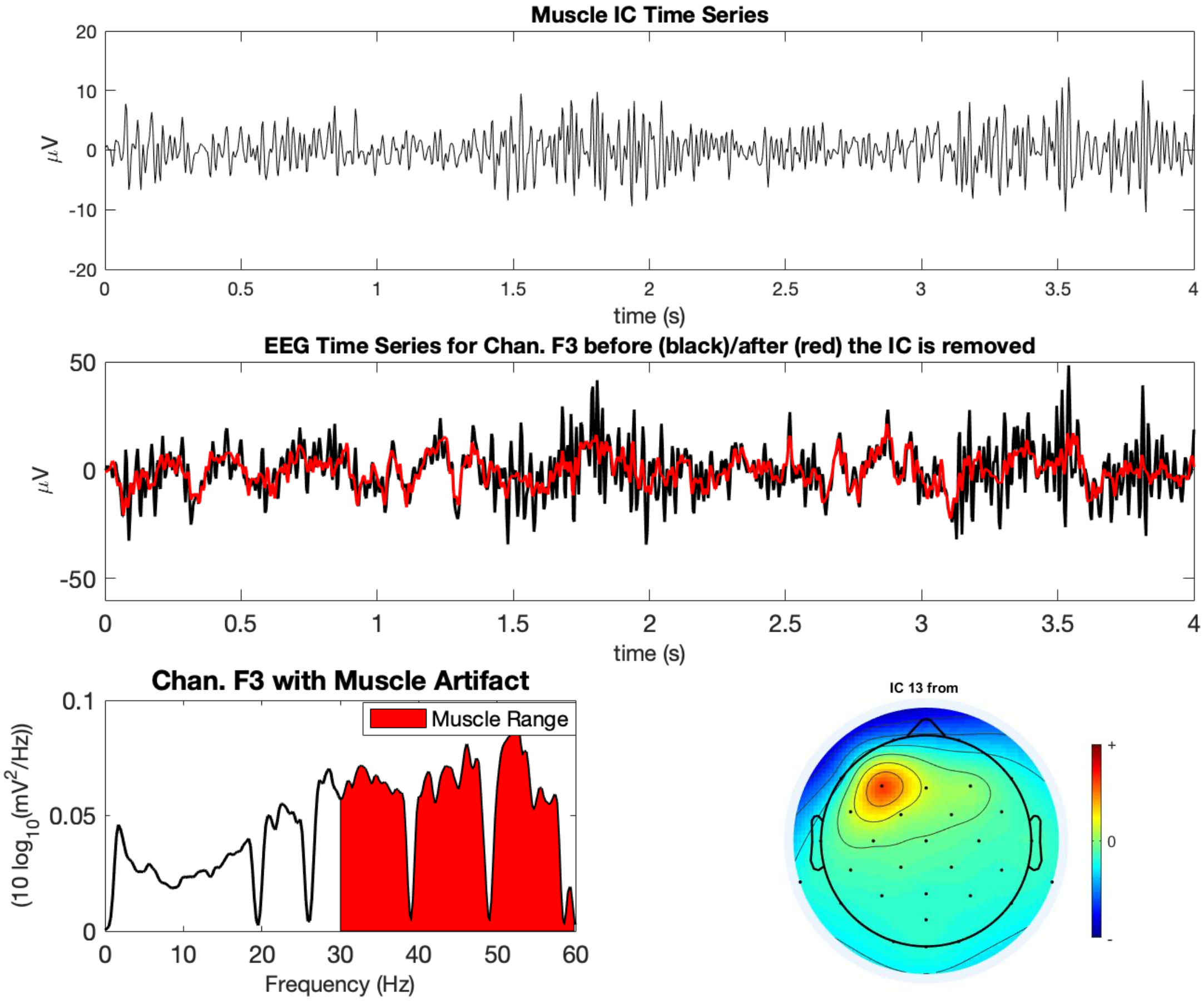

**Supplementary Figure S5:** An example of a Muscle Component. Muscle electrical activity or "electromyogenic" (EMG) artifacts exhibit widespread high-frequency activity due to asynchronous motor action units. These components are flagged if the power of the signal is spread out in frequencies higher than 30 Hz, known as the gamma band

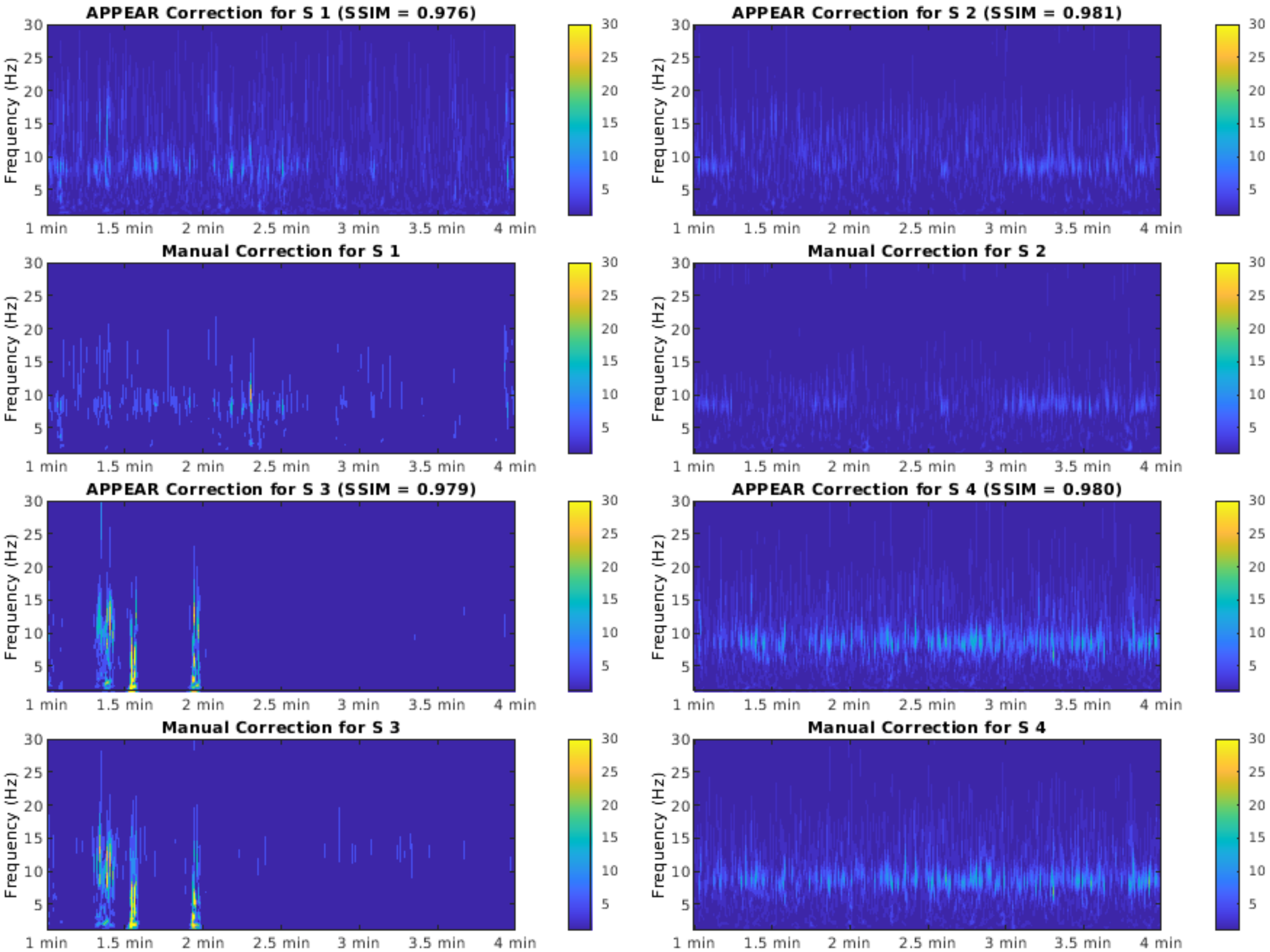

**Supplementary Figure S6:** A comparison between CWT results from APPEAR and manually corrected data for subjects 1 through 4.

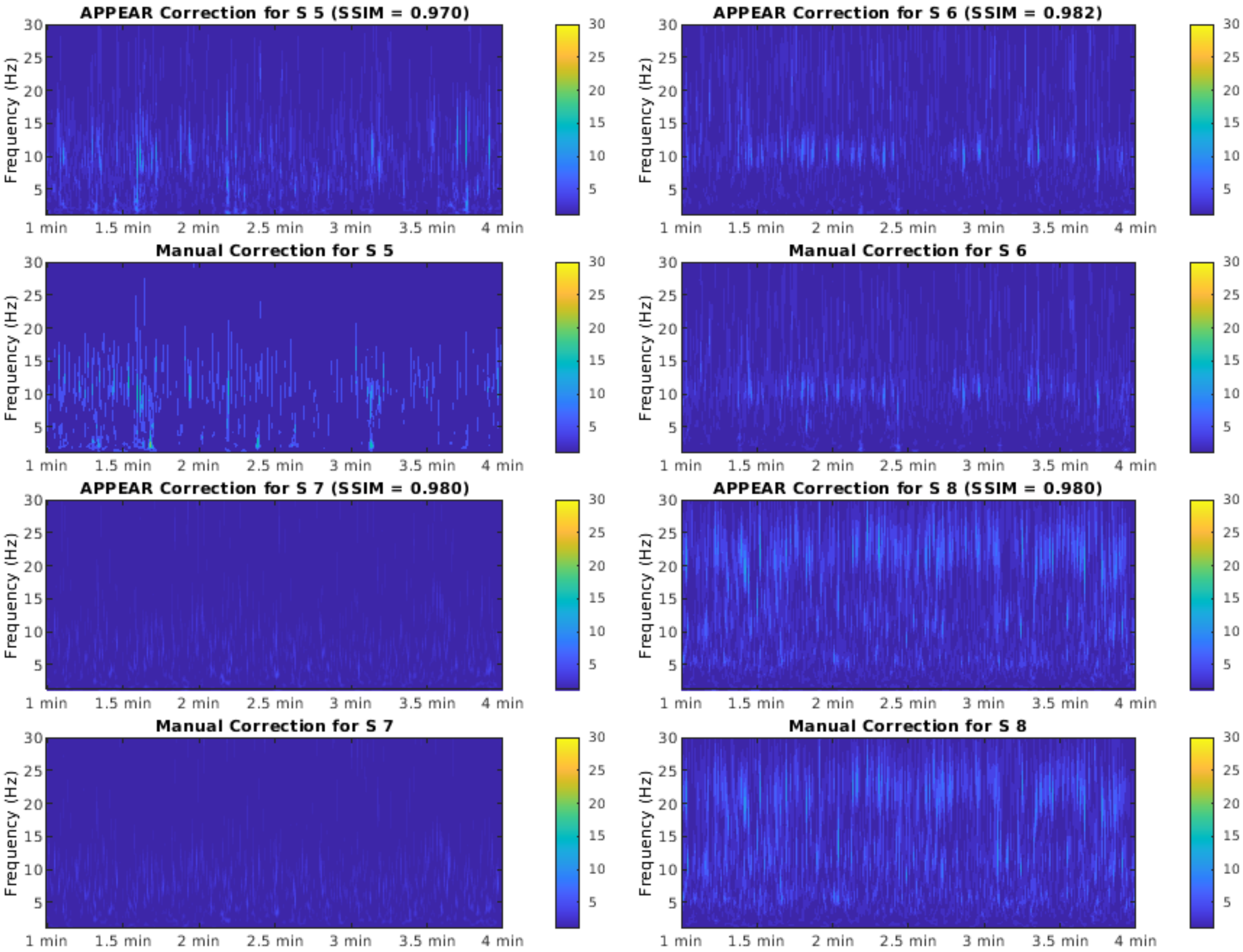

**Supplementary Figure S7:** A comparison between CWT results from APPEAR and manually corrected data for subjects 5 through 8.

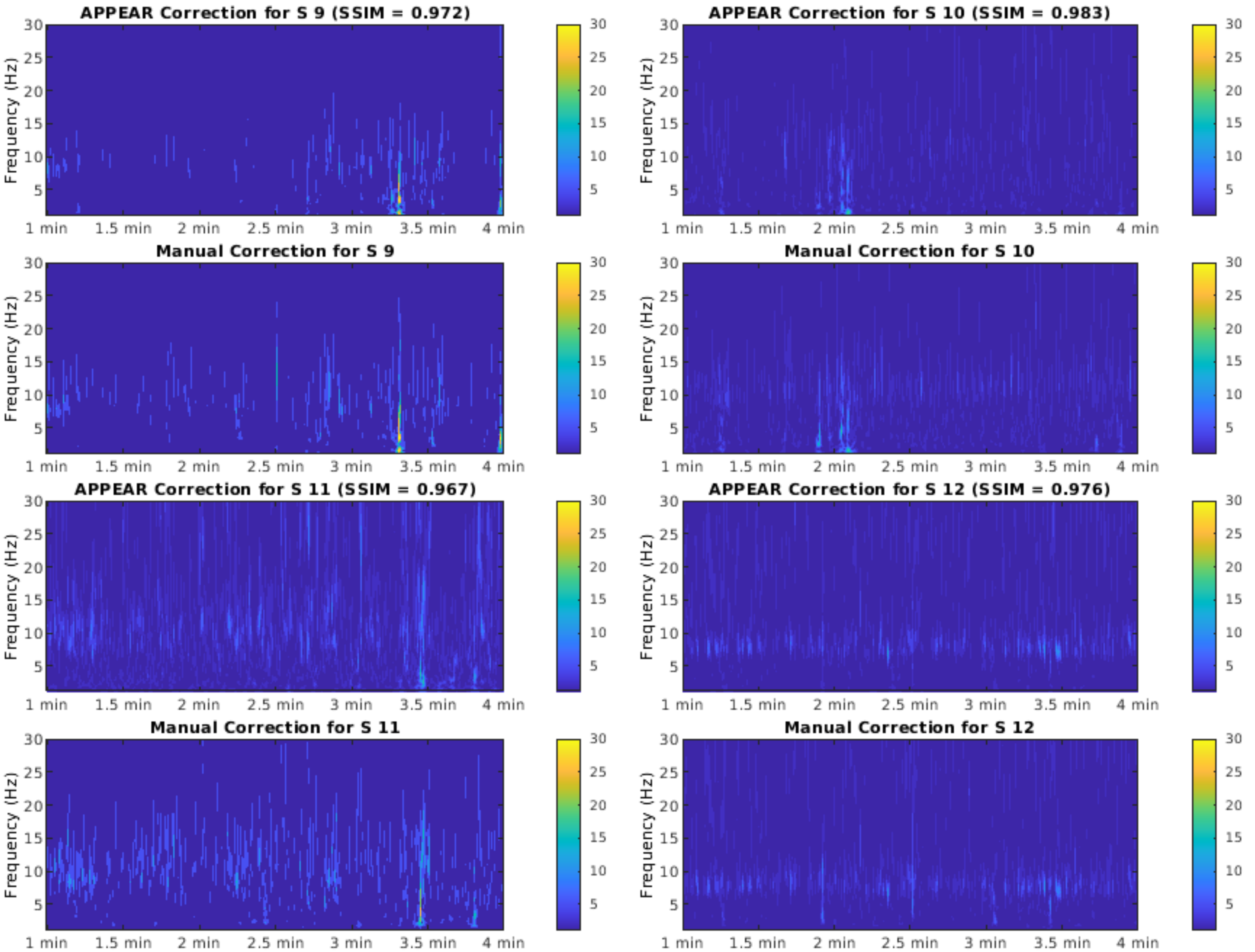

**Supplementary Figure S8:** A comparison between CWT results from APPEAR and manually corrected data for subjects 9 through 12.

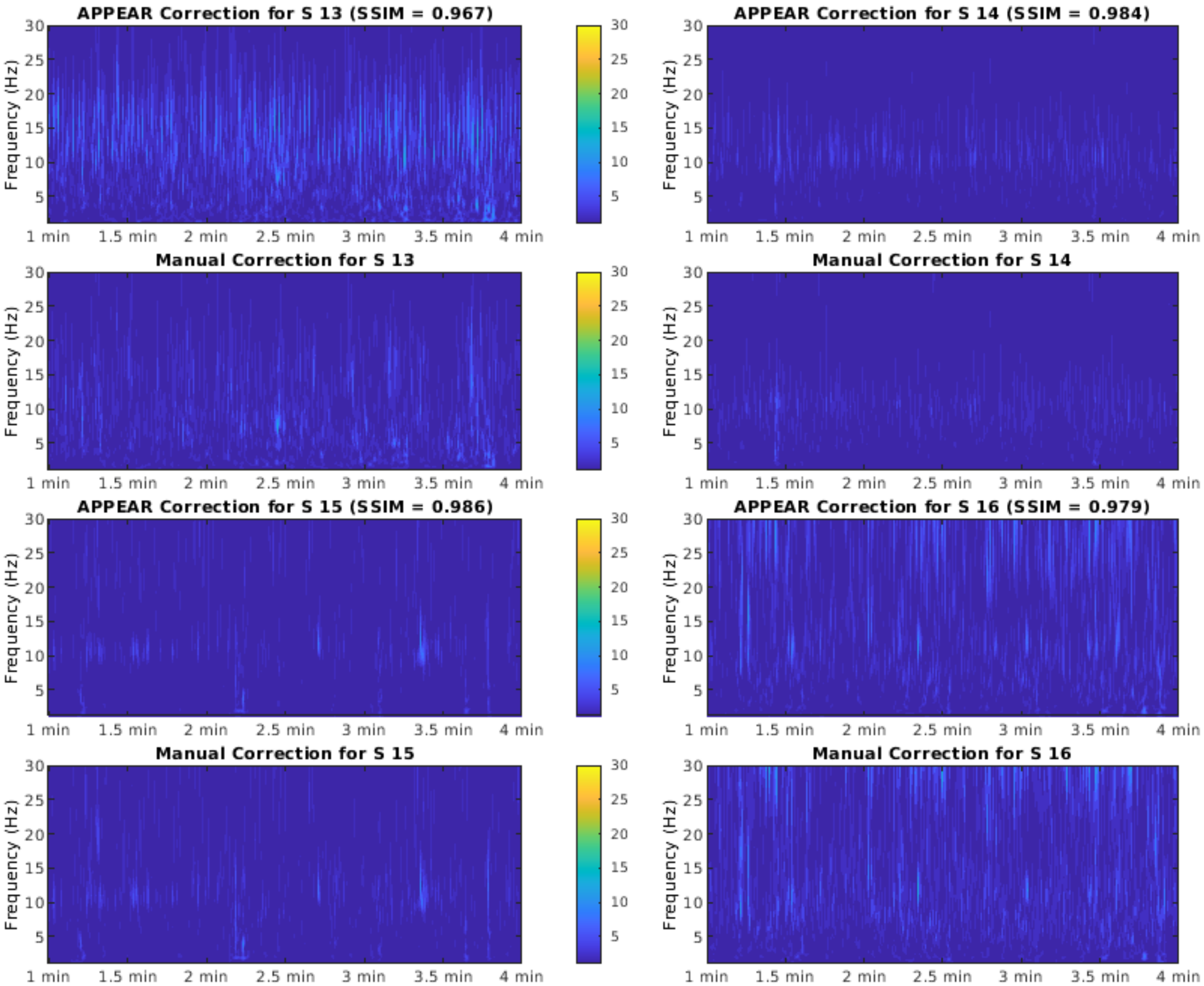

**Supplementary Figure S9:** A comparison between CWT results from APPEAR and manually corrected data for subjects 13 through 16.

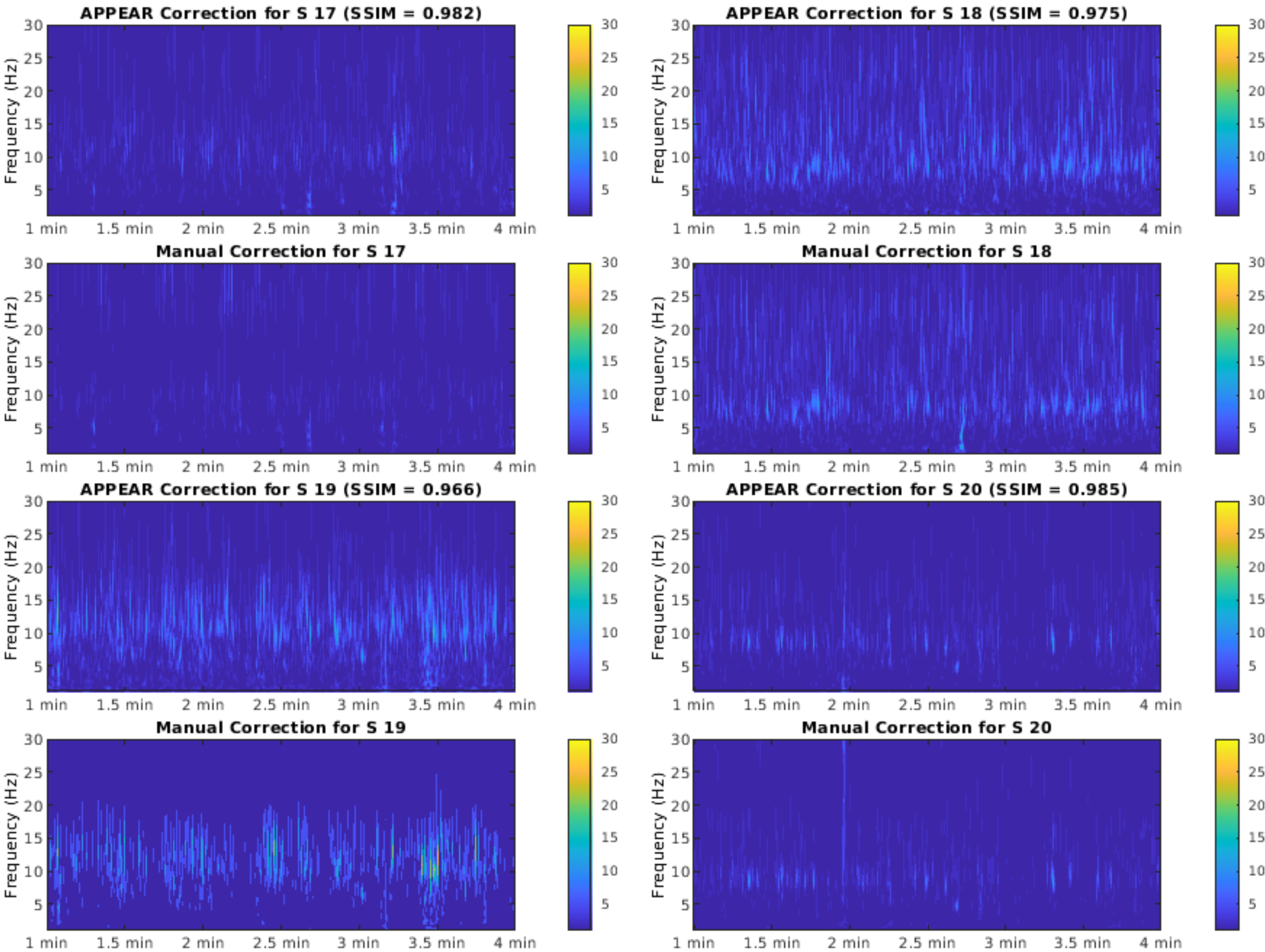

**Supplementary Figure S10:** A comparison between CWT results from APPEAR and manually corrected data for subjects 17 through 20.

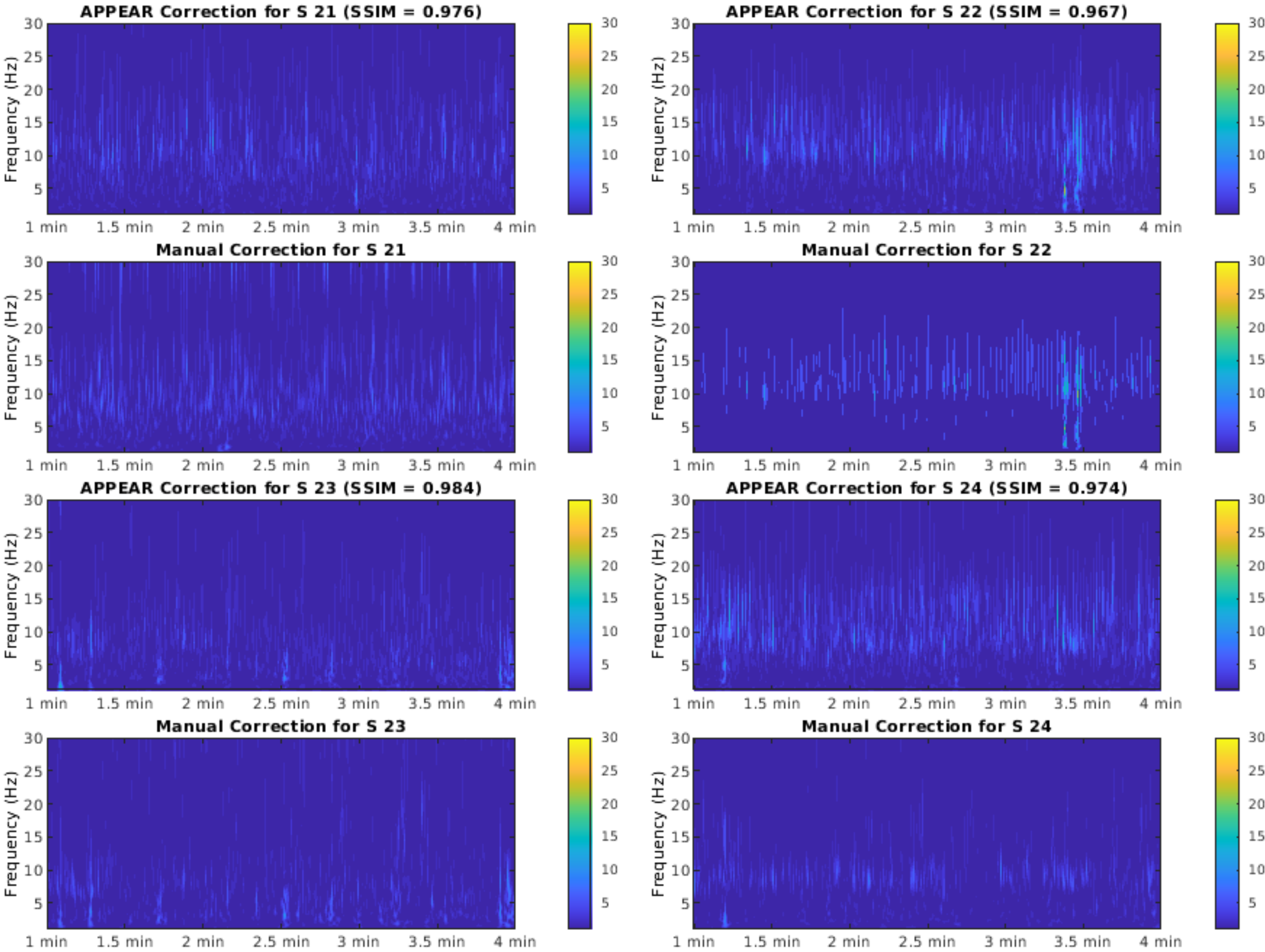

**Supplementary Figure S11:** A comparison between CWT results from APPEAR and manually corrected data for subjects 21 through 24.

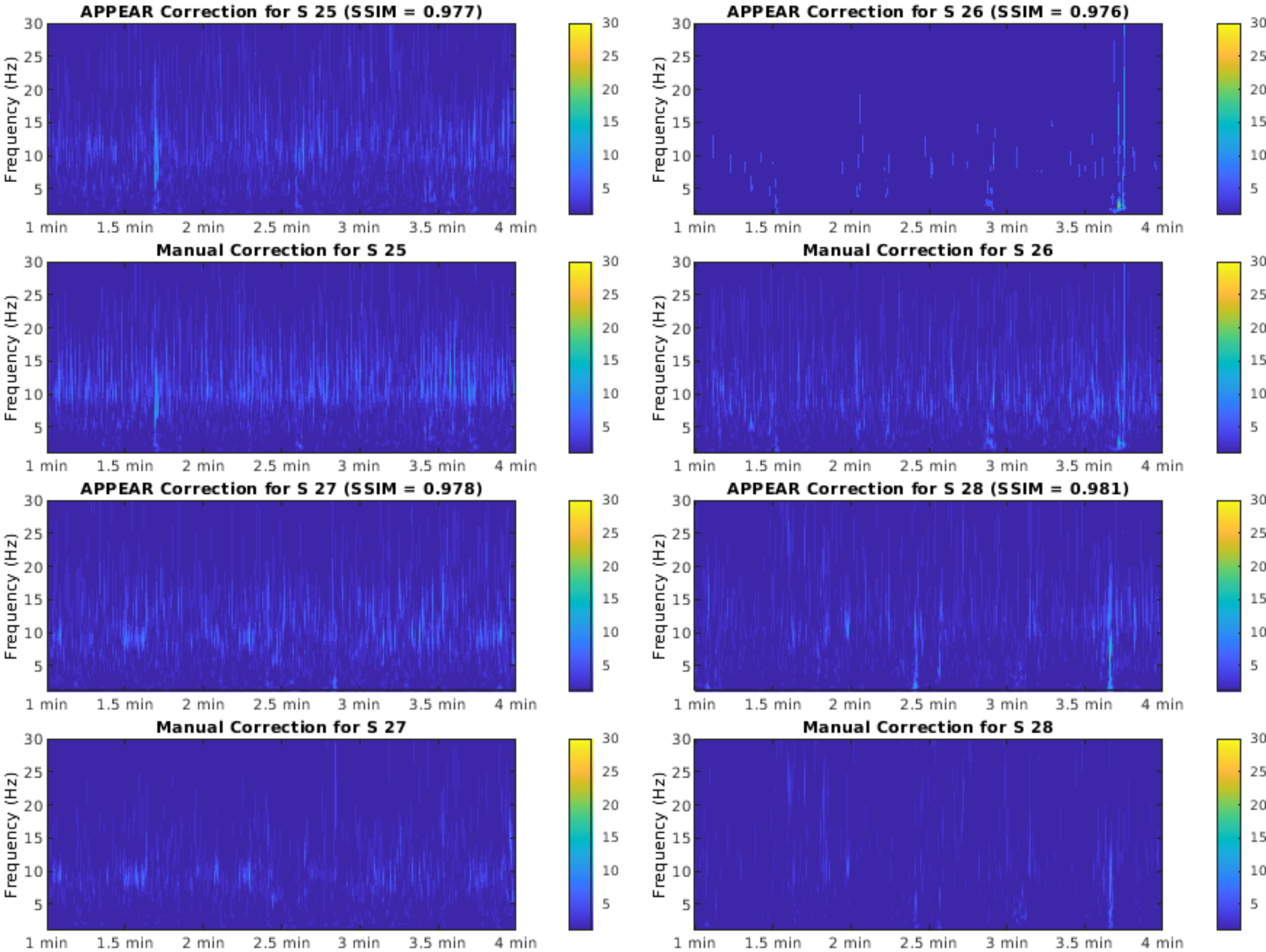

**Supplementary Figure S12:** A comparison between CWT results from APPEAR and manually corrected data for subjects 25 through 28.

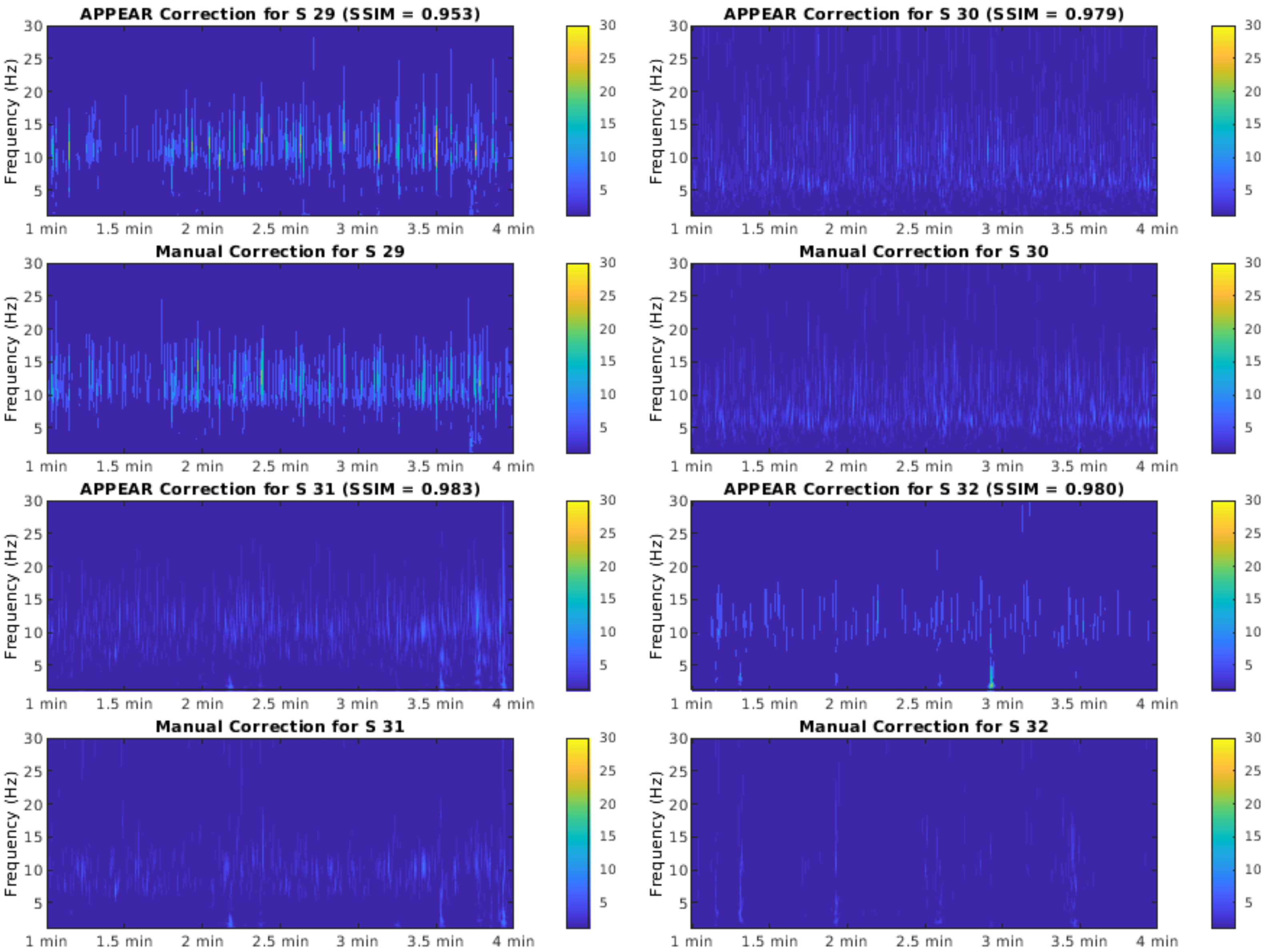

**Supplementary Figure S13:** A comparison between CWT results from APPEAR and manually corrected data for subjects 29 through 32.

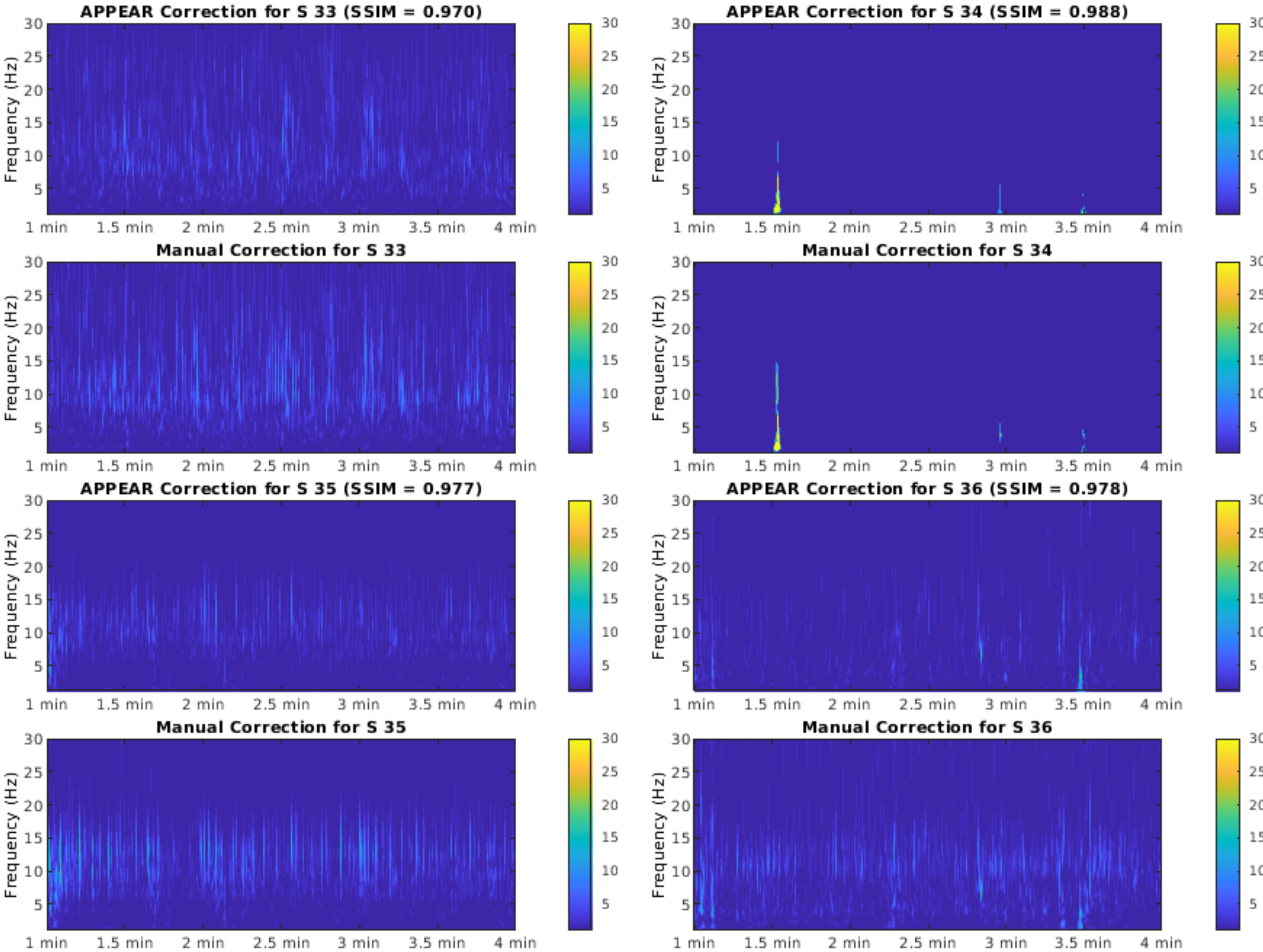

**Supplementary Figure S14:** A comparison between CWT results from APPEAR and manually corrected data for subjects 33 through 36.

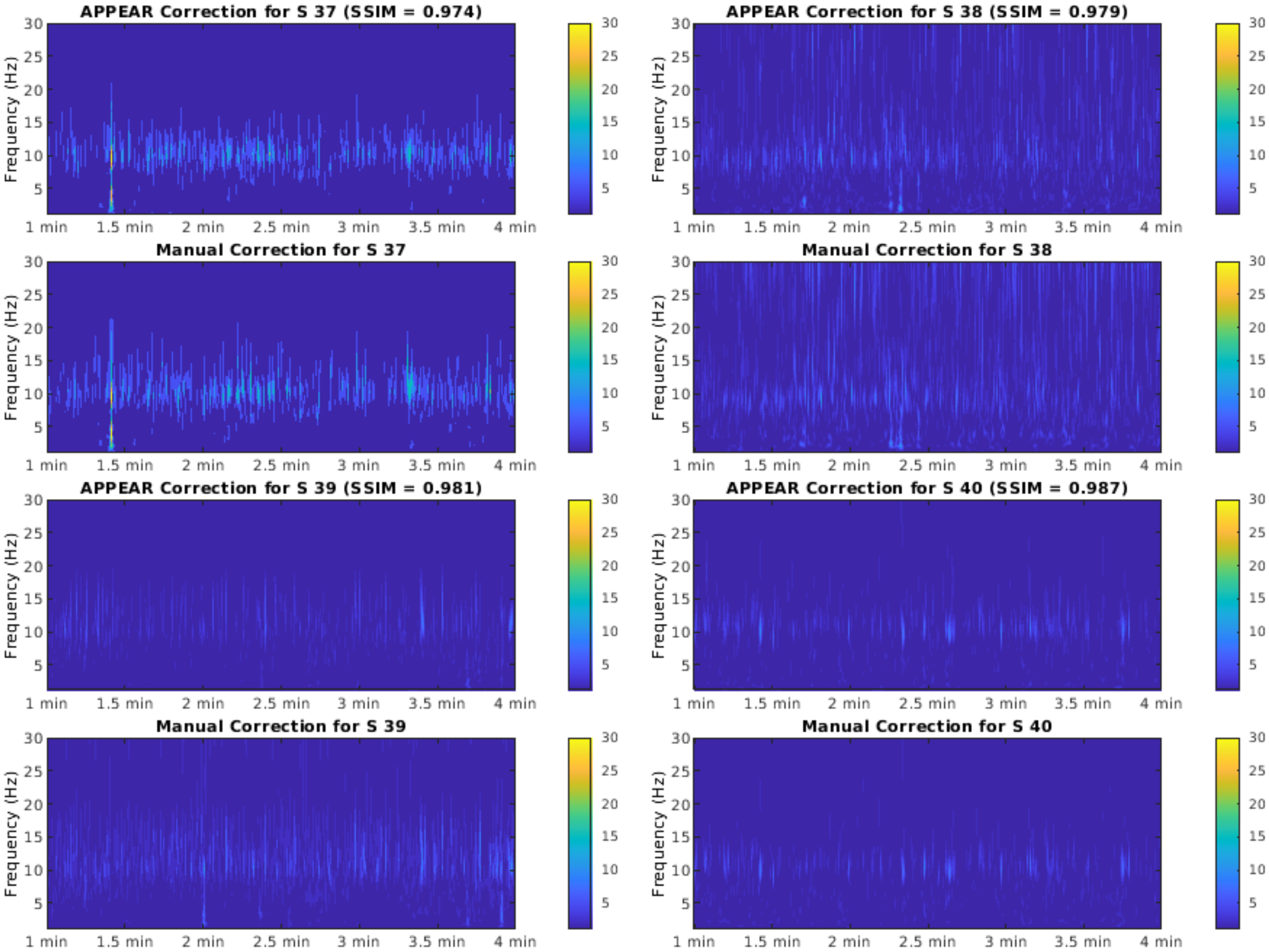

**Supplementary Figure S15:** A comparison between CWT results from APPEAR and manually corrected data for subjects 37 through 40.

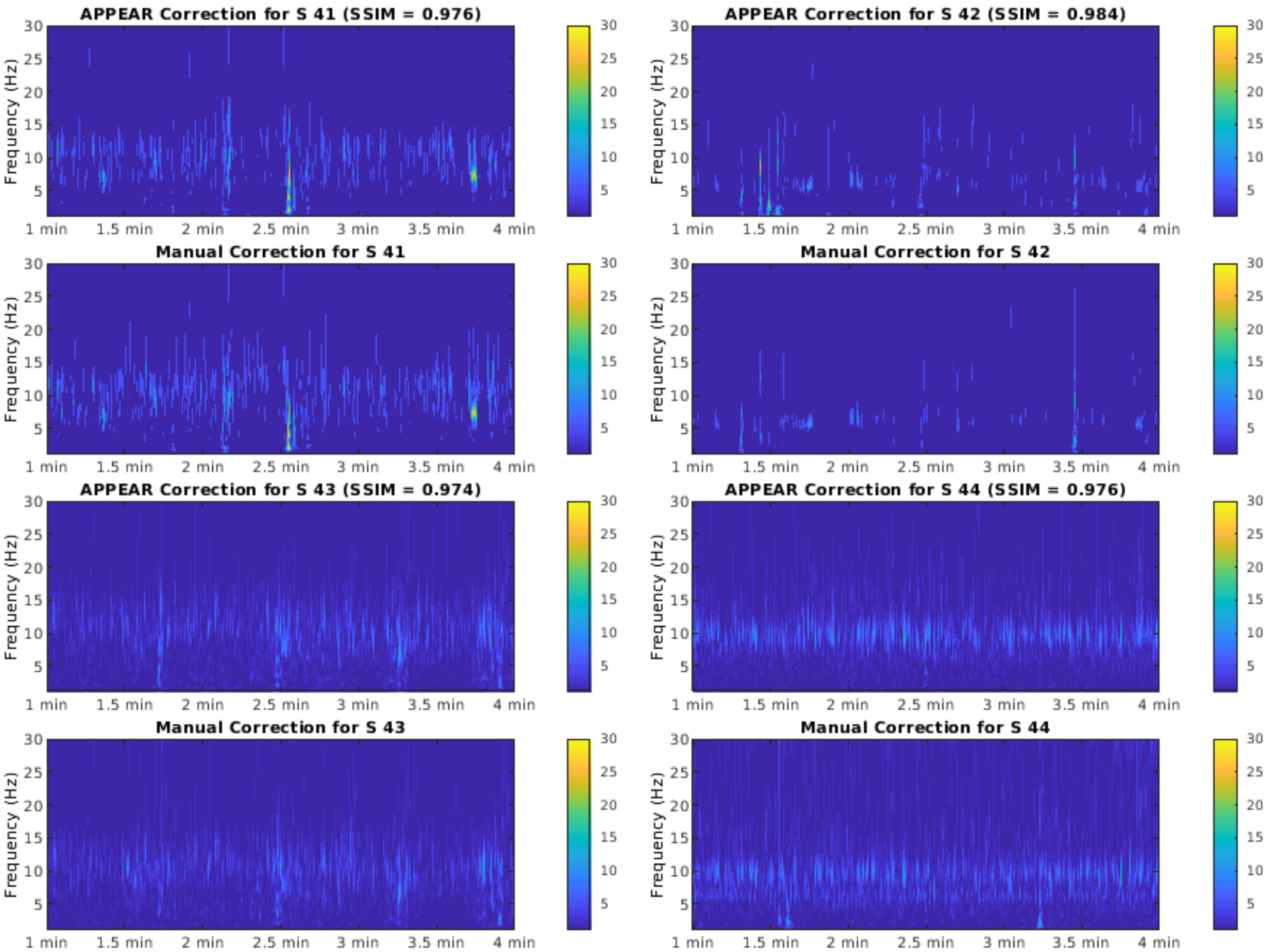

**Supplementary Figure S16:** A comparison between CWT results from APPEAR and manually corrected data for subjects 41 through 44.

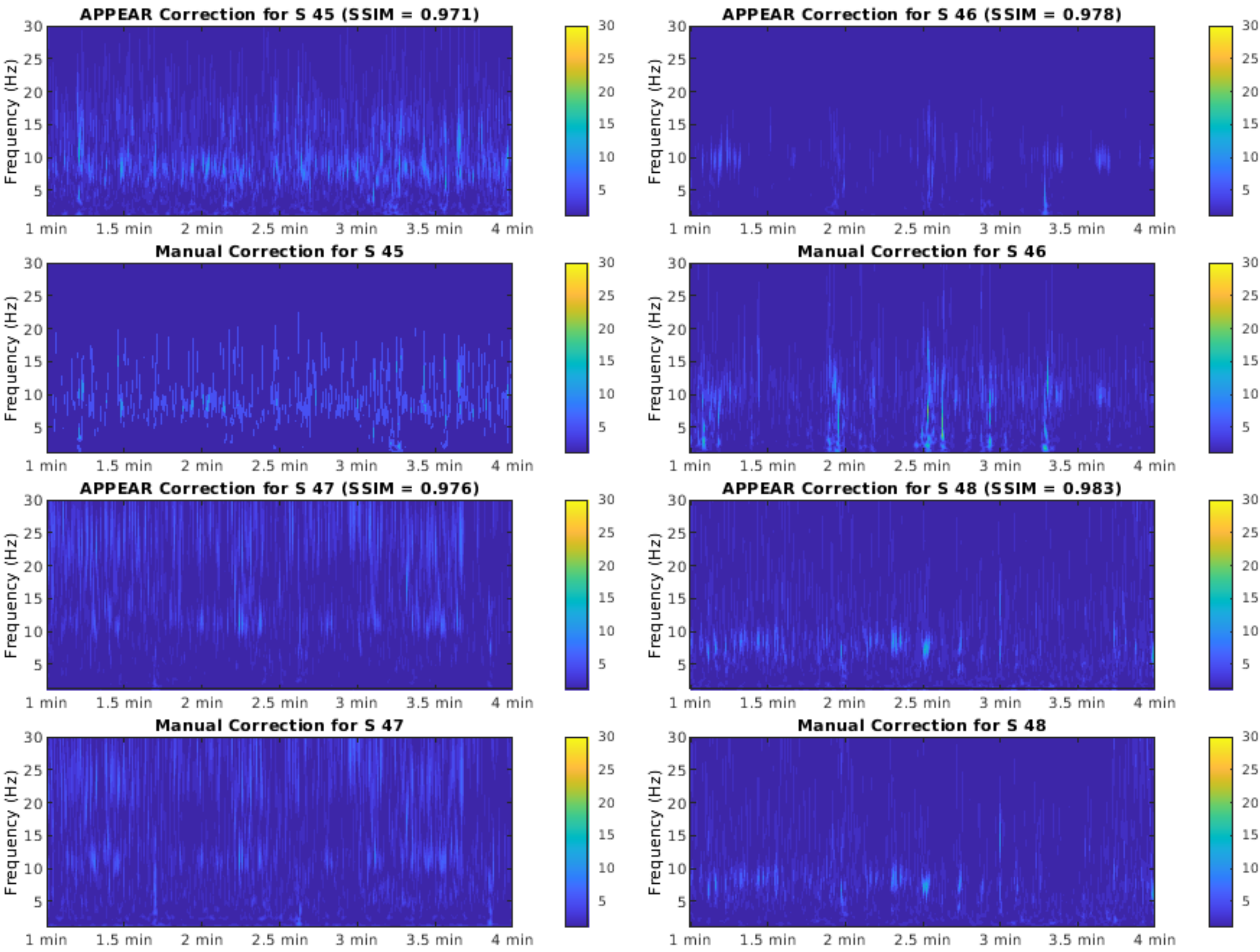

**Supplementary Figure S17:** A comparison between CWT results from APPEAR and manually corrected data for subjects 45 through 48.

**Supplementary Table S1.** The average heart rate for all subjects (Rest and Stop Signal experiments) derived from: i) the simultaneous ECG signal recorded with EEG (calculated with the FMRIB plug-in implemented in MATLAB); ii) EEG data (computed using independent component analysis); iii) physiological pulse oximetry signal concurrently and independently recorded during EEG-fMRI (MATLAB peak detection function - findpeaks). The values in parentheses show the absolute heart rate difference measured between ECG/EEG and pulse oximetry.

| **Rest** | **Subject1** | **Subject2** | **Subject3** | **Subject4** | **Subject5** | **Subject6** | **Subject7** | **Subject8** |
|---|---|---|---|---|---|---|---|---|
| ECG (FMRIB) | 62.79 (0.38) | 62.86 (0.42) | 66.27 (0.19) | 72.62 (0.09) | 69.59 (0.02) | 79.46 (0.18) | 76.00 (0.05) | 44.98 (0.08) |
| EEG (ICA-Based) | 63.18 (0.02) | 117.83 (54.55) | 66.44 (0.02) | 72.56 (0.04) | 69.53 (0.07) | 79.56 (0.27) | 76.03 (0.03) | 45.03 (0.03) |
| Pulse oximetry (peak detection) | 63.17 | 63.28 | 66.46 | 72.52 | 69.61 | 79.29 | 76.06 | 45.06 |
| **Stop Signal** | | | | | | | | |
| ECG (FMRIB) | 66.58 (0.09) | 68.93 (4.36) | 65.44 (0.92) | 75.80 (0.06) | 76.99 (0.09) | 85.01 (0.523) | 77.31 (2.040) | 43.86 (0.17) |
| EEG (ICA-Based) | 66.07 (0.41) | 64.45 (0.12) | 64.49 (0.02) | 76.33 (0.47) | 76.87 (0.03) | 85.08 (0.60) | 75.46 (0.19) | 44.00 (0.02) |
| Pulse oximetry (peak detection) | 66.49 | 64.57 | 64.52 | 75.86 | 76.90 | 84.49 | 75.27 | 44.03 |